\documentclass[final,3p,times]{elsarticle}

\usepackage[T1]{fontenc}
\usepackage[utf8]{inputenc}
\usepackage{enumerate}
\usepackage{rotating}
\usepackage{lscape}
\usepackage{pifont}
\usepackage{tikz}
\usepackage{multirow}
\usepackage{lineno,hyperref}
\modulolinenumbers[5]

\newcommand{\tabincell}[2]{
	\begin{tabular}{@{}#1@{}}#2\end{tabular}
}

\usepackage{acronym}
\acrodef{ML}[ML]{Machine Learning}
\acrodef{DL}[DL]{Deep Learning}
\acrodef{ADE}[ADE]{Automatic Depression Estimation}
\acrodef{HAMD}[HAMD]{Hamilton Rating Scale for Depression}
\acrodef{BDI}[BDI]{Beck Depression Index}
\acrodef{DCNN}[DCNN]{Deep Convolutional Neural Network }

\begin{document}

\begin{frontmatter}

\title{Deep Learning for Depression Recognition with Audiovisual Cues: A Review}


\author[mymainaddress,mysecondaryaddress]{Lang He\corref{mycorrespondingauthor}}
\ead{langhe@xupt.edu.cn}
\author[mythirdaryaddress,myfouthaddress]{Mingyue Niu}
\author[aalto]{Prayag Tiwari\corref{mycorrespondingauthor}}
\ead{prayag.tiwari@aalto.fi}
\author[aalto]{Pekka Marttinen\corref{mycorrespondingauthor}}
\ead{pekka.marttinen@aalto.fi}
\author[NWU]{Rui Su\corref{mycorrespondingauthor}}
\ead{sabrina@nwu.edu.cn}
\author[jjw]{Jiewei Jiang\corref{mycorrespondingauthor}}
\ead{jiangjw924@126.com}
\author[chgg]{Chenguang Guo\corref{mycorrespondingauthor}}
\ead{guochg@nwpu.edu.cn}
\author[mymainaddress,mysecondaryaddress]{Hongyu Wang}
\author[mymainaddress,mysecondaryaddress]{Songtao Ding} 
\author[mymainaddress,mysecondaryaddress]{Zhongmin Wang}
\author[mymainaddress,mysecondaryaddress]{Xiaoying Pan}
\author[myfiveaddress]{Wei Dang}
\cortext[mycorrespondingauthor]{Corresponding author}

\address[mymainaddress]{School of Computer Science and Technology, Xi'an University of Posts and Telecommunications, Xi'an Shaanxi 710121, China}
\address[mysecondaryaddress]{Shaanxi Key Laboratory of Network Data Analysis and Intelligent Processing, Xi'an University of Posts and Telecommunications, Xi'an Shaanxi 710121, China}
\address[mythirdaryaddress]{National Laboratory of Pattern Recognition (NLPR), Institute of Automatic Chinese Academy of Sciences (CASIA), Beijing 100190, China}
\address[myfouthaddress]{School of Artificial Intelligence, University of Chinese Academy of Sciences (UCAS), Beijing 100049, China}
\address[aalto]{Department of Computer Science, Aalto University, Espoo, Finland.}

\address[NWU]{School of Foreign Languages, Northwest University, Xi'an Shaanxi, China}
\address[jjw]{School of Electronic Engineering, Xi’an University of Posts and Telecommunications, Xi’an, China}
\address[chgg]{School of Electronics and Information, Northwestern Polytechnical University, Xi’an Shaanxi, China}
\address[myfiveaddress]{Shaanxi Mental Health Center, Xi'an Shaanxi 710061, China}

\begin{abstract}
With the acceleration of the pace of work and life, people have to face more and more pressure, which increases the possibility of suffering from depression. However, many patients may fail to get a timely diagnosis due to the serious imbalance in the doctor-patient ratio in the world. Promisingly, physiological and psychological studies have indicated some differences in speech and facial expression between patients with depression and healthy individuals. Consequently, to improve current medical care, many scholars have used deep learning to extract a representation of depression cues in audio and video for automatic depression detection. To sort out and summarize these works, this review introduces the databases and describes objective markers for automatic depression estimation (ADE). Furthermore, we review the deep learning methods for automatic depression detection to extract the representation of depression from audio and video. Finally, this paper discusses challenges and promising directions related to automatic diagnosing of depression using deep learning technologies.

\end{abstract}

\begin{keyword}
Affective computing, Depression, Deep learning, Automatic depression estimation, Review
\end{keyword}

\end{frontmatter}


%

\section{Introduction}\label{sec:Introduction}

Depression is a type of mental disorder, which brings serious burden to individuals, families, and society. According to the World Health Organization (WHO), depression will be the most common mental disorder by 2030 \cite{mathers2006projections}. In severe situations, depression leads to suicide \cite{kessler2003epidemiology,hawton2013risk}. Besides,
the report released by \cite{hawton2013risk,mcgirr2007examination} points out that approximately 50\% of suicides are linked to depression.  Currently, there is no unique and efficient clinical characterization of depression, which makes the diagnosis of depression time-consuming and subjective \cite{Maj2020TheCC}. As gold-standard assessments or tools mainly depend on the subjective experience of clinicians, it is challenging to have a unified standard for diagnosing the severity of depression. The main diagnostic tools for severity, e.g., \ac{HAMD} \cite{hamilton1986hamilton}, rely on interviews conducted by clinicians or individuals themselves, yielding a score which summarizes the behavior of the patients. Diagnosing depression is complicated, depending not only on the educational background, cognitive ability, and honesty of the subject to describe the symptoms, but also on the experience and motivation of the clinicians. Comprehensive information and thorough clinical training are needed that to diagnose the severity of depression accurately \cite{mundt2007voice}. 
Some biological markers, for instance, low serotonin levels \cite{nock2008suicide,sharp20115}, neurotransmitter dysfunction \cite{luscher2011gabaergic,poulter2008gabaa} and genetic abnormalities \cite{dwivedi2003altered, gatt2009interactions}, have been considered as indicators of depression, however it is unclear which biomarker(s) is the most efficient as an indicator. Hence, in recent years, numerous \ac{ADE} systems have been introduced to automatically estimate the severity scale of depression based on audiovisual cues extracted with techniques developed in machine learning, speech recognition, and computer vision field  \cite{cohn2009detecting,cummins2011investigation,joshi2013multimodal,scherer2013automatic}.
  
\begin{figure}[h]
	\centering
	\centerline{\includegraphics[scale=0.75]{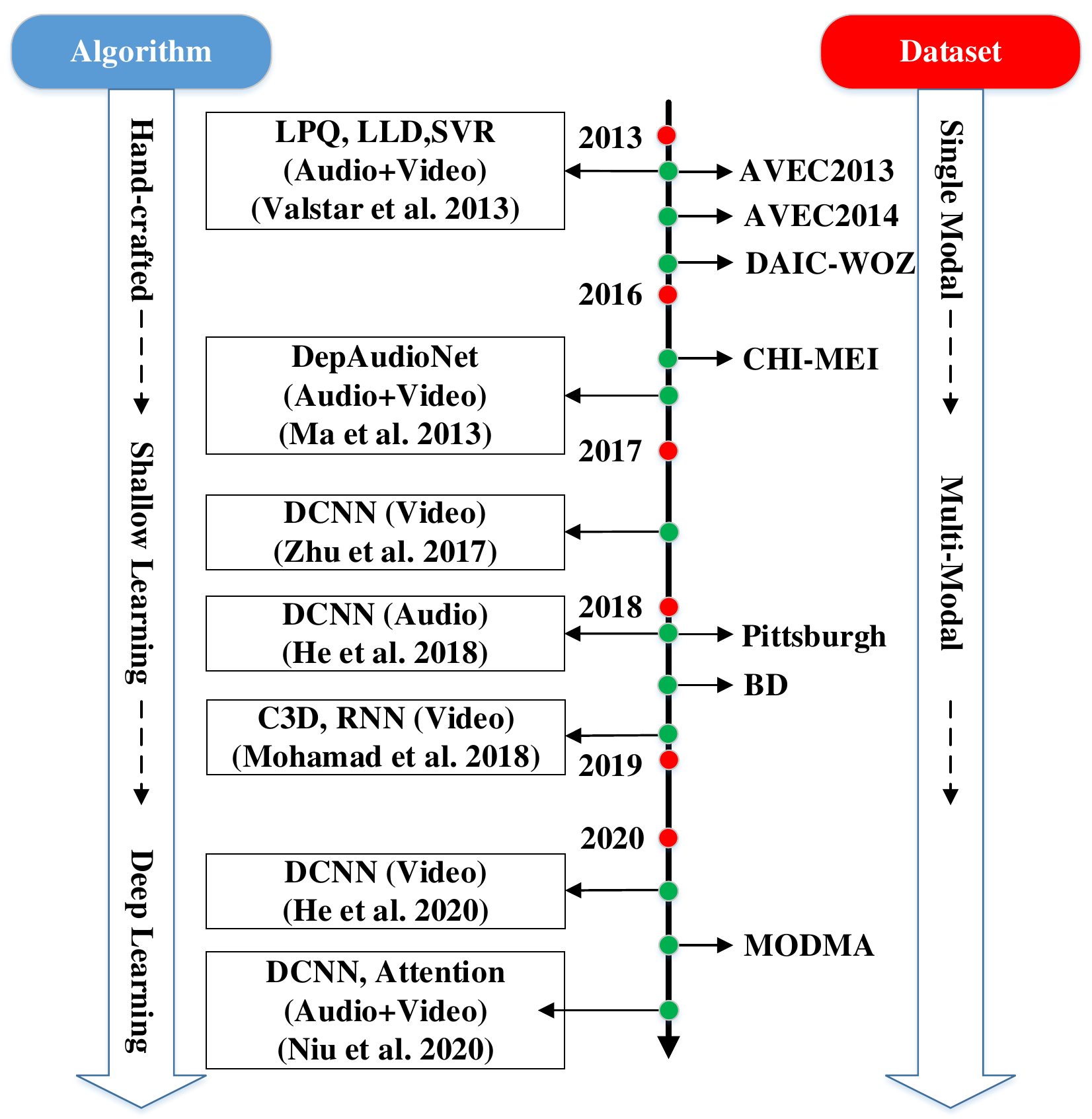}}
	\caption{A brief chart to show the evolution of \ac{ADE} with approaches and databases. From 2013 to 2021, the algorithm based on feature extraction had undergone three different stages from hand-crafted to shallow to deep learning. In the meantime the databases have evolved from single-modal (e.g., audio or video alone) to multi-modal (comprising multiple data types jointly). }
	\label{fig:Time}	
\end{figure}

Designing a representative feature and its extraction for predicting the scale (i.e., severity) of depression is an important step in the deep learning architecture for \ac{ADE}. \ac{ADE} features can either be hand-crafted or based on deep learning models. Examples of widely used hand-crafted features include Local Binary Patterns (LBP) \cite{shan2009facial}, Local Phase Quantization from Three Orthogonal Planes (LPQ-TOP) \cite{wen2015automated}, Local Binary Patterns from Three Orthogonal Planes (LBP-TOP) \cite{zhao2007dynamic}, and others (e.g., Facial Action Units (FAUs), Landmarks, Head Poses, Gazes) \cite{du2019encoding}. However, since 2013, depression recognition challenges such as Audio-Visual Emotion Recognition Challenge (AVEC2013) \cite{valstar2013avec} have recorded depression data by Human-Computer Interaction. Meanwhile, the fast development of deep learning has motivated many scientists to study \ac{DL} approaches for depression recognition, which has resulted in a promising performance compared to the hand-crafted features. For deep learning approaches, a wide range of studies have adopted the Deep Convolutional Neural Network (DCNN) to extract multi-scale feature representations \cite{ma2016depaudionet,jan2017artificial,song2018human,uddin2020depression,al2018video,zhu2017automated,de2019combining,dedepression,song2020spectral,Md2020Depression,he2018automated}. Fig. \ref{fig:Time} shows this evolution of \ac{ADE} according to methods and databases. 

\subsection{Limitations}

In recent years, some exhaustive reviews on depression recognition and analysis have been published based on audio \cite{cummins2015review}, and visual cues \cite{pampouchidou2017automatic}. These surveys provided a comprehensive scope for \ac{ADE}. Yet, there remain two unexplored aspects. As existing reviews only focus either on audio or visual cues for estimating the depression scale, the combined use of audiovisual cues has not been adequately discussed. In addition, existing surveys only consider traditional approaches, and \ac{DL} technology has not yet been covered in their analyses. Recently, \ac{DL} technology has quickened the development and innovation of depression recognition based on audiovisual cues. As far as we know, an in-depth review on multi-modal audiovisual approaches for depression recognition is still missing. Our goal is to fill the gap in the existing extensive reviews by including the increasingly important deep multi-modal \ac{ADE} approaches, based on audiovisual cues.

Despite the superior performance of deep learning, it still has some issues for \ac{ADE}. First, training of deep learning methods (e.g. DCNN, Recurrent Neural Network (RNN), Convolutional 3D (C3D)) needs a lot of data and the existing public depression databases are limited and inadequate for the depression detection task. Second, individuals have different personal attributes, e.g., gender, age, race, educational background, etc. These personal attributes and traits have not been accounted for in existing deep learning models. For different individuals, these attributes allow considering depression from different perspectives. For instance, the education of an individual may indicate a broad interest in a lot of things, which may maybe preventive from depression. Third, combining hand-crafted and deep learning based features is a challenge that needs to be addressed to guarantee excellent performance in depression detection. Fourth, a fusion of multi-modal signals requires a solid theoretical foundation to mine the complementary representations from the different modalities.  

\begin{figure}[h]
	\centering
	\centerline{\includegraphics[scale=0.6]{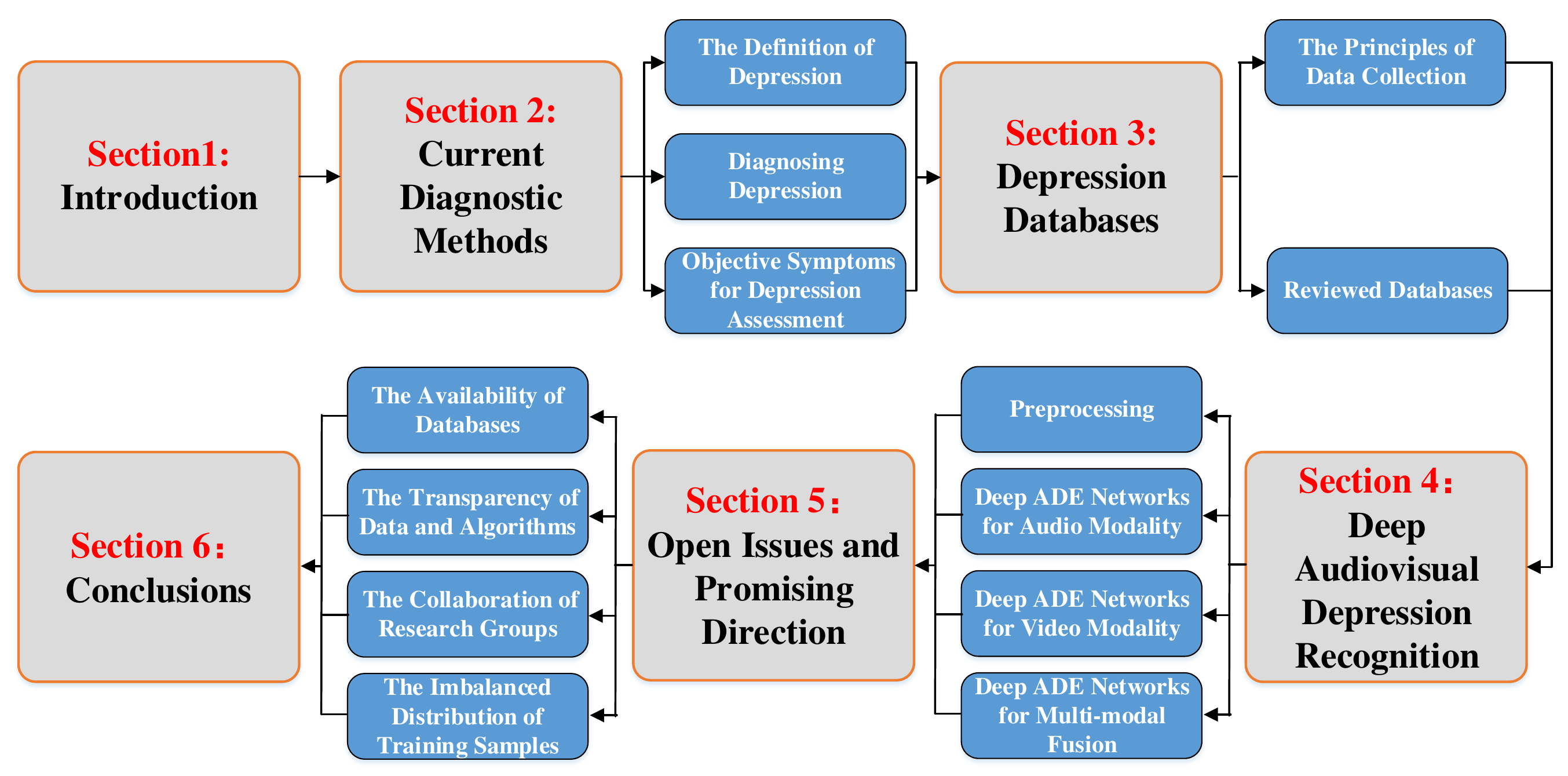}}
	\caption{The paper structure through graphical illustration.}
	\label{fig:section_pipeline_crop}	
\end{figure}

In this paper, we comprehensively review automatic depression detection methods based on deep neural networks, discuss the challenges, and point to future research directions. In the following, Section \ref{sec:related} provides the definition of depression and describes the objective markers for depression assessment. Section \ref{sec:Databases} introduces several  multi-modal depression databases. Section \ref{sec:methods} gives a detailed review of general deep \ac{ADE} methods and presents several novel neural network architectures based on audiovisual cues. Additional issues are described in Section \ref{sec:Additional}. Section \ref{sec:Conclusions} provides conclusions based on our exposition. In addition, to make a clear explanation, the structure of the paper is shown in Fig. \ref{fig:section_pipeline_crop}.

\section{Current Diagnostic Methods}\label{sec:related}    

To better understand the procedure of depression recognition based on audiovisual cues, the definition of depression is reviewed next, and then automatic methods to diagnose depression are surveyed.

\subsection{The Definition of Depression}\label{sec:definition}

\begin{figure}[h]
	\centering
	\includegraphics[scale=1.7]{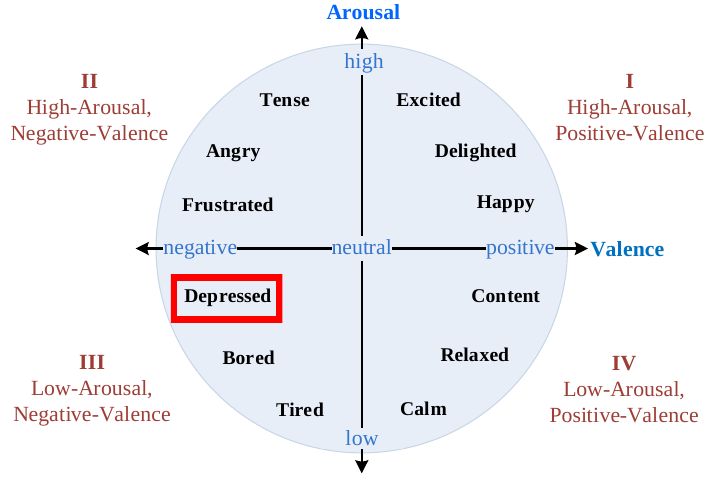}
	\caption{The two-dimension emotion space, which can be divided into four quadrants \cite{yu2016building}. Every quadrant is associated with various emotions. For instance, high-arousal and positive-valence emotional states include excited, delighted, happy, etc. The depression disorder, which can be categorized in the third quadrant of the VA.} 
	\label{fig:framework}
\end{figure}

In 1980, Russell \cite{russell1980circumplex} proposes that emotional states can be represented as continuous numerical vectors in a two-dimensional space, called the  Valence-Arousal (VA) space, see Fig. \ref{fig:framework}. The valence dimension refers to the two types of emotional states, i.e., positive and negative. The arousal dimension represents the intensity of emotion from sleepiness (or boredom) to high excitement. As shown in Fig. \ref{fig:framework}, depression is located in the third quadrant of the VA space \cite{russell1980circumplex}, which corresponds to low-arousal and negative valence. 

According to the definition from the Diagnostic and Statistical Manual of Mental Disorders (DSM) of the American Psychiatric Association (APA) \cite{american2013diagnostic}, depression can be further divided as follows: Major Depressive Disorder (MDD), Persistent Depressive Disorder (Dysthymia), Disruptive Mood Dysregulation Disorder (DMDD), Premenstrual Dysphoric Disorder (PDD), Substance/Medication-Induced Depressive Disorder (S/M-IDD), Depressive Disorder Due to Another Medical Condition (DDDAMC), and Other Specified Depressive Disorder (OSDD) or Unspecified Depressive Disorder (UDD). The DSM has provided the general criteria for classifying mental disorders by observed symptoms. When an individual has at least one of the following symptoms: 1) depressed mood most of the day and/or 2) markedly diminished interest or pleasure, in combination with at least four or more of the symptoms in Table \ref{table:symptoms} that have  sustained for at least two weeks. In addition, the aforementioned symptoms are also expected to cause clinically significant distress or impairment in social, occupational, or other important areas of functioning. Nonetheless, these different types of depression related disorders manifest themselves in a similar way to a certain extent. 
\begin{table*}[h]
	\centering
	\caption{Symptoms associated with depression \cite{american2013diagnostic}.}
	\label{table:symptoms}
	\begin{tabular}{c}
		\hline
		\leftline{\textbf{Depressed Mood and/or Markedly diminished interest or pleasure}} \\
		\hline
		\leftline{\textbf{In combination with four of:}} \\
		\parbox{.9\linewidth}{\begin{enumerate}
				\item Significant weight loss when not dieting OR weight gain (e.g., a change of  more than 5\% of body weight in a month), decrease OR increase in appetite  nearly every day. In children, a failure to make expected weight gains
				\item Insomnia OR hypersomnia nearly every day (inability to sleep OR excessive sleeping)
				\item Psychomotor agitation OR retardation nearly every day (observable by others, not merely subjective feelings of restlessness OR of being slowed down)
				\item Feelings of worthlessness OR excessive OR inappropriate guilt (which may be delusional) nearly every day (not merely self-reproach OR guilt about being sick)
				\item Diminished ability to think OR concentrate, OR indecisiveness, nearly every day (either by subjective account OR as observed by others)
				\item Fatigue OR loss of energy almost every day
				\item Recurrent thoughts of death (not just fear of dying), OR recurrent suicidal ideation without a specific plan, a suicide attempt OR a specific plan for committing suicide
		\end{enumerate}}\\
		\hline
	\end{tabular}
\end{table*}

The question of how to diagnose depression has attracted attention of many researchers from different fields. However, the understanding of the pathogenesis of depression has not yet been unified and agreed upon. Nevertheless, pathogenesis is usually considered to be linked with a dysfunction of the cortical-limbic system, reducing its activity and connectivity \cite{deckersbach2006functional,evans2006using,mayberg2005deep,niemiec2006alpha}. Nonetheless, it is believed that depression depends on the interaction between a genetic predisposition and environmental factors \cite{nestler2002neurobiology,Cadoret1990EarlyLP}. For the effect of genetic predisposition, in \cite{lesch2004gene}, the authors found that monkeys suffered from depression may deprived of their mother. For the effect of environmental factors, in \cite{Cadoret1985GeneticAE}, Remi et al. found that for males, drinking too much in an adoptive family increases the risk of depression. For females, the death of an adoptive parent prior to adoptee age of 19 or the presence of an individual with a behavioural disorder in the adoptive family increases the risk of depression.

In addition, DSM has often been criticized that the boundaries between mental illnesses are not always properly defined. This resulted in subjective biases \cite{brown2001reliability,kamphuis2009categorical,lux2010deconstructing,oquendo2008issues,stein2010mental,watson2005rethinking}. There exist at the fewest 1497 unique profiles for depression \cite{ostergaard2011heterogeneity}. In some cases of the same diagnosis, two depressed subjects may not share any identical symptom \cite{jh2012verbal}. It is often considered that MDD can be referred to as \textit{Clinical Depression}. As reported in the works of \cite{chow2019h} and \cite{sobocki2006cost}, the incremental economic loss of MDD has grown from 2005 to 2010 by 21.5\% in the United States, while the economic loss is evaluated at 1\% of the GDP. 

\subsection{Diagnosing Depression}\label{sec:Automatic}

It is difficult in primary care settings to assess the depression severity. Diagnosis of depression is often complicated with the chance of misidentification, the time-consuming nature, and the fact that not all depressed subjects directly show depressive manifestations (e.g., helplessness or hopelessness, etc.) \cite{mitchell2009clinical, schumann2012physicians}. In addition, the combination of biological elements, family/environmental stressors, and personal vulnerabilities plays a vital role in affecting the onset of MDD \cite{kessler2013epidemiology}.    

Currently, the most frequently used assessment methods are interviews, e.g., \ac{HAMD} \cite{hamilton1986hamilton}, or self-assesments, e.g., the \ac{BDI} (the first edition in 1961, and the recent version of 1996) \cite{beck1996comparison}. A score is assigned by the assessment methodologies (\ac{HAMD} and \ac{BDI}) to each patient for characterizing their severity level by rating the 21 depression related symptoms. The main difference between \ac{HAMD} and \ac{BDI} is that \ac{HAMD} requires a 20-30 min interview where the clinician fills the rated questionnaire, while \ac{BDI} needs 5-10 min to complete the self-reported questionnaire. In addition, the two rating scales consider difference measures; the \ac{HAMD} concentrates on neuro-vegetative symptoms (e.g., psychomotor retardation, weight, sleep, and fatigue, etc.), while the \ac{BDI} focuses on self-assessments of negative self-evaluation symptoms. \ac{HAMD} and \ac{BDI} have been proven to obtain consistency when making a distinction between depressed from non-depressed patients \cite{baer2010handbook, maust2012psychiatric}. The \ac{HAMD} tool was considered a gold standard for diagnosing the severity of depression. However, related research has exposed some issues \cite{maust2012psychiatric, bagby2004hamilton, gibbons1993exactly}. Most importantly, some typical symptoms (i.e., insomnia, low mood, agitation, anxiety, and reduced weight) related to the severity of depression are neglected by the HAMD. For every question in the \ac{HAMD} questionnaire, the psychologists or clinicians should provide 3-5 possible responses to rate the severity of depression. A score in one of the ranges 0-2, 0-3, and 0-4 is assigned to indicate the severity of each symptom of depression. The summed score can be divided into five groups: \textit{(Normal: the range from 0 to 7)}, \textit{(Mild: the range from 8 to 13)}, \textit{(Moderate: the range from 14 to 18)}, \textit{(Severe: the range from 19 to 22)} and \textit{(Very Severe, $\geq$23)}. While \ac{HAMD} has covered numerous symptoms of depression, \cite{bech1981hamilton,faries2000responsiveness} commented that only a part of the listed symptoms was useful for estimating the severity of depression. Nevertheless, a simple “symptom checklist” approach is considered insufficient to assess the \ac{ADE}.

As mentioned above, the definition of depression from the clinical perspective may also depend on the scores provided by Self-report Scales and Inventories (SRSIs). The common assessment tools are \ac{BDI}/\ac{BDI}-II, PHQ-2/8/9 (Patient Health Questionnaire, 2, 8, or 9 is the number of questions, respectively), and Depression and Somatic Symptoms Scale (DSSS). To obtain a further comprehension of SRSI, \ac{BDI} is introduced as follows. \ac{BDI} is a commonly used assessment tool for SRSI of depression \cite{cusin2009rating}. It consists of $21$ questions, including cognitive, affective, and somatic symptoms, and several negative manifestations (e.g., self-evaluations and self-criticisms) related to depression. Every item of \ac{BDI}/\ac{BDI}-II, which is defined by multiple-choice options, is weighted by a numerical value (range: $0$ -- $3$). The range of \ac{BDI} scores is from $0$ to $63$ (\textit{(no or minimal depression: the range from 0 to 13)}, \textit{(Mild: the range from 14 to 19)}, \textit{(Moderate: the range from 20 to 28)}, \textit{(Severe: the range from 29 to 63)}). Originally, \ac{BDI} was not designed specifically for primary care usage, but its practical performance \cite{nuevo2009usefulness} shows that it is valid also for meetings in the primary care. 

Though SRSIs have been widely used in various studies with the specificity and sensitivity reaching up to 80\% to 90\%, they nonetheless pose certain problems \cite{williams2005performance}. Specifically, the SRSI does not consider the clinical meaning of the observed symptoms and it allows individual variability when reporting different traits or characteristics, in contrast to a clinical interview \cite{pichot1986self}. Furthermore, SRSI cannot differentiate very well among different depression subtypes \cite{baer2010handbook}. In addition, SRSI is susceptible to intentional or unintentional reporting bias \cite{ben2003assessing}. Overall, despite the problem in  providing an efficient diagnosis of depression \cite{gilbody2008screening, ren2015performance, stockings2015symptom}, SRSIs have been widely adopted in various ways, for example in the primary health care, and research. Some researchers emphasized the cost-effectiveness SRSIs as a way for widespread screening practices to promote depression assessment \cite{mitchell2009screening}. 

Table \ref{table: commonly used scale of depression} lists some depression scale ratings, e.g., \ac{HAMD}, \ac{BDI}, PHQ-9, Inventory of Depressive Symptomology (IDS), the 16-item Quick Inventory of Depressive Symptomology (QIDS), Zung Self-Report Depression Scale (Zung-SDS) and 10-item Montgomery–A Sberg Depression Rating Scale (MADRS), etc.   

\begin{table*}[!t]
	\centering
	\setlength{\tabcolsep}{0.8mm}
	\caption{Commonly utilized depression rating scales \cite{maust2012psychiatric}.}
	\begin{tabular}{ccccc}
		\hline		
		{\bf Scale}& {\bf Interview} & {\bf Self Assessment} & {\bf  {\tabincell{c}{Number \\ of items}}} & {\bf {\tabincell{c}{Time to \\ complete (Seconds)}}} \\
		\hline		
		\ac{HAMD} \cite{hamilton1986hamilton} & $\surd$ &  & 17 or 21 & 20-30  \\
		\ac{BDI} \cite{beck1996comparison}    &  & $\surd$ & 21 & 5-10  \\
		PHQ \cite{kroenke2002phq} &  & $\surd$ & 9 &<5  \\		
		QIDS \cite{Rush2003The} &  & $\surd$ & 16 & 5-10 \\
		MARSD \cite{montgomery1979new}  & $\surd$  &  & 10 & 20-30  \\
		{\tabincell{c} {Inventory of Depressive \\
				Symptomology (IDS) \cite{rush1996inventory}}}                     & $\surd$ &  & 30 & 10-15                     \\
		{\tabincell{c}  {Zung Self-Report Depression \\
				Scale (Zung-SDS)\cite{zung1965self}}}                      &  & $\surd$ & 20 & 5-10 \\   
		\hline			                   
	\end{tabular}
	\label{table: commonly used scale of depression}
\end{table*}

\subsection{Objective Makers for Depression Assessment}\label{sec:bio}

It is generally considered that the representation of depression can be affected by various aspects \cite{ellgring1989non,waxer1974therapist}. Observable behavioral signals were not accepted in the psychiatry field. However, several studies in these fields have still obtained popularity up to the present. Objective markers have been widely adopted in psychology, which can be utilized as an objective diagnostic tool in related fields (i.e., primary clinical settings, psychological institutions). It offered a robust assessment tool for assisting the clinicians in diagnosing the severity scale effectively and offered later feedback and valuable advice for susceptible individuals. With the development of wearable devices, an interactive virtual tool is designed to deploy on the smartphone platform to help diagnose the depression subjects or the suspected people \cite{scherer2013automatic}. Therefore, there is an urgent requirement for designing novel assessment tools, such as developing diagnostic tools to investigate new markers. Previous studies on objective physiological, biological, and behavioral markers have improved the efficiency of psychiatric diagnosis and, they have the latent capacity to decrease the socio-economic costs caused by depression \cite{jh2012verbal,costanza2014neurobiology}.

In the early works of Emil Kraepelin \cite{kraepelin1921manic}, who was recognized as the father of modern psychiatry, he defined depressed voice as ``patients speak in a low voice, slowly, hesitatingly, monotonously, sometimes stuttering, whispering, try several times before they bring out a word, become mute in the middle of a sentence''. In \cite{cummins2015review}, the speech was considered a key objective maker for depression analysis, covering a wide range of features (i.e., prosodic, source, acoustic, and vocal tract dynamic). 

In addition, patterns around facial regions were also significant for depression estimation. Hands and body posture are included in certain patterns related to depression assessment. Visual cues are essential for depression detection. It is considered that pupil dilation has a relation with depression. In \cite{siegle2011remission}, the authors consider that faster pupillary represents positive by healthy controls. Depressed subjects represent slower pupil dilation responses in certain conditions \cite{silk2007pupillary,jones2015motivational,wang2014pupillometry,zhou2015tackling,kudinova2016pupillary,li2016alleviated}. In \cite{price2016anxious}, the authors found that pupil bias and diameter are also important for assessing depression symptoms. In addition, the factors of facial expressions (e.g., anger, sadness, joy, surprise, disgust, fear, etc.) were regarded as a discriminative cue for depression detection. Suppose an individual is diagnosed to have depression symptoms. In that case, they will indicate low expressibility when showing the facial expressions\cite{ellgring1989non,stratou2013automatic,ghosh2014multimodal,stratou2015automatic,scherer2013automatic,yu2013multimodal,morency2015simsensei,zhou2015tackling}. The features consist of reduced eye contact \cite{lucas2015towards}, gaze direction \cite{scherer2013audiovisual,gratch2014distress,scherer2014automatic,morency2015simsensei}, eyelid activity \cite{6738869}, iris movement \cite{alghowinem2015cross}, and eye openings/blinking \cite{zhou2015tackling,alghowinem2015cross,gupta2014multimodal}. Moreover, Eye movement and blinking were also considered a discriminative feature for predicting depression \cite{6738869}. Furthermore, the duration of spontaneous smiles \cite{scherer2013audiovisual, scherer2014automatic}, the intensity of smile \cite{scherer2013audiovisual,morency2015simsensei,scherer2014automatic}, mouth animation \cite{gupta2014multimodal}, the lack of smiles \cite{lucas2015towards} were also considered that contain valuable patterns for depression detection. 

Action Units (AU) are the basic actions of muscle groups or individual muscles, which was originally proposed by Ekman et al. \cite{ekman2002facial}, and then adopted by Cohn et al. \cite{cohn2009detecting} to analyze the depression state. Meanwhile, a new AU-based method named Region Units (RU) was proposed \cite{mcintyre2009approach}.  Region Units (RU) are used to represent the regions of the face that enclose AUs. Various works have adopted AU to estimate the scale of depression and obtained promising performances \cite{stratou2013automatic,ghosh2014multimodal,stratou2015automatic,mcintyre2010computer,cohn2010social,mcintyre2011facial,girard2014nonverbal,girard2013social,cohn2013beyond,williamson2014vocal,mandal2014understanding,yang2017coupled,valstar2016avec,yang2016decision,yang2017hybrid,yang2018integrating}. Later, it was found \cite{6738869} that head pose and movement also contained discriminative patterns for assessing the severity of depression \cite{stratou2013automatic,ghosh2014multimodal,stratou2015automatic,scherer2013automatic,scherer2013audiovisual,zhou2015tackling,girard2014nonverbal,yu2013multimodal,morency2015simsensei,scherer2014automatic,alghowinem2015cross,gupta2014multimodal,6681444,joshi2012depression,joshi2013automated,joshi2013can}. Specifically, there were 46 points used to train a 3D face model using Active Appearance Models (AAM) to extract the head pose and movement features. Additionally, facial animation and the variability of motor were also utilized for depression detection \cite{stratou2013automatic,stratou2015automatic,scherer2013automatic,gupta2014multimodal}. Body posture (i.e., the upper part of the body, the lower part of the body, and hands) are also very significant feature for detection \cite{joshi2012depression,joshi2013automated,joshi2013can,joshi2012neural,joshi2013relative}. In \cite{gratch2014distress}, the researchers considered that foot tapping and self-adapters also influence depression detection. In addition, the activity from facial muscles, skin electrical reactions, and peripheral blood pressure also caused involuntary changes, which often reflect the common and persistent negative thoughts and sad emotions of depression. In \cite{hosseinifard2013classifying}, they found that electroencephalographic recordings may have certain patterns related to depression. In \cite{adorni2016could} and \cite{ho2020diagnostic}, Functional Near-Infrared Spectroscopy (fNIRS) is also considered as a tool to assist in the depression diagnosis task \cite{suto2004multichannel}. 

In addition, several works have indicted that depression can cause the change of neurophysiological and neurocognitive abnormalities, which are demonstrated from an individual communication via facial gesturing, vocal articulation and so on \cite{scherer2015self,pampouchidou2017automatic}. Therefore, we will focus on audiovisual signals for \ac{ADE} in this review.

\begin{sidewaystable*}[h]
	\centering
	\caption{Summary of the Audiovisual databases which have been adopted in the reviewed works for the last 20 years.      Abbreviations:
		DPRD – Depressed, SCDL – Suicidal, NTRL – Neutral, not depressed or suicidal, M – Number of males, F – Number of Females DSM - Diagnostic and
		Statistical Manual of Mental Disorders, \ac{HAMD} - Hamilton Rating Scale for Depression, \ac{BDI} -Beck Depression Inventory, QIDS - Quick Inventory of
		Depressive Symptomology, PHQ-9 - Patient Health Questionnaire. Note: where DSM is present as a clinical score for all depressed patients in corpus to meet criteria for Major Depressive Disorder}
	\setlength{\tabcolsep}{1.5mm}	
	\begin{tabular}{cccccc}	
		\hline
		\multicolumn{1}{c}{\bf Database} & \multicolumn{1}{c}{\bf Modality} &
		\multicolumn{1}{c}{\bf \tabincell{c}{Subjects}} &		
		\multicolumn{1}{c}{\bf Annotation} & \multicolumn{1}{c}{\bf Ground Truth}  &\multicolumn{1}{c}{\bf Public /Private}\\
		\hline
		\textcircled{1}: DementiaBank \cite{becker1994natural}(1994) & A+V+T  &226&  \ac{HAMD}  &Clinical
		Assessment&Public \\	
		\textcircled{2}: -- \cite{stassen1998speech}(1998)&A&43&\ac{HAMD}  (DPRS = \ac{HAMD} $\geq
		$ 10 )&--&Private\\
		\textcircled{3}: -- \cite{france2000acoustical}(2000)&A&115&{\tabincell{c}{DSM-IV, \ac{BDI} \\ (DPRS = \ac{BDI} > 20)}}&Clinical
		Assessment&Private\\
		\textcircled{4}: -- \cite{alpert2001reflections}(2001)&A&41&DSM-\uppercase\expandafter{\romannumeral3}-R, \ac{HAMD} (DPRS = \ac{HAMD} $\geq
		$ 18)&Clinical Assessment&Private\\
		\textcircled{5}: -- \cite{moore2004comparing}(2004)&A&33&DSM-IV&Clinical
		Assessment&Private\\
		\textcircled{6}: -- \cite{yingthawornsuk2006objective}(2006)&A&32&{\tabincell{c}{\ac{BDI} \\ (DPRD = \ac{BDI} > 20)}}&{\tabincell{c}{Interviews with\\ a Therapist}}&Private\\
		\textcircled{7}: -- \cite{cohn2009detecting}(2009)&A&57&{\tabincell{c}{DSM-IV, \ac{HAMD} \\ (DPRD = \ac{HAMD} $\geq
				$ 15)}}&--&Private\\
		\textcircled{8}: ORI \cite{maddage2009video}(2009)	&   V                     &139&   Manual annotation                     &--      &Private                 \\
		\textcircled{9}: BlackDog \cite{Alghowinem2012From}(2009)	&    A+V                    &80&    DSM-IV, $HAMD\textgreater15 $        &Clinical
		Assessment &Private                       \\
		\textcircled{10}: ORYGEN \cite{ooi2011prediction}(2011)	&   V                    &191&  Manual annotation                      & Clinical
		Assessment    &Private                 \\
		\textcircled{11}: -- \cite{mundt2012vocal} (2012)&Audio&165&{\tabincell{c}{\ac{HAMD}, DSM-IV,\\QIDS-C, QIDS (QIDS-SR)}}&Clinical assessment&Private\\				
		\textcircled{12}: AVEC2013 \cite{valstar2013avec} (2013)&    A+V  &292& \ac{BDI}-II &Self-report &Public                       \\
		\textcircled{13}: AVEC2014 \cite{valstar2014avec} (2014) &   A+V  &292&  \ac{BDI}-II  &Self-report &Public             \\
		\textcircled{14}: Crisis Text Line \cite{chen2014visualizations} (2014) &  T &--& Manual annotation                       & --  &Private                       \\
		\textcircled{15}: DAIC-WoZ \cite{gratch2014distress} (2014)&  A+V+T                      &110& {\tabincell{c}{PHQ-9 \\(DPRD = PHQ-9 > 10) }}                     & Self-report&Public                     \\
		\textcircled{16}: Rochester \cite{zhou2015tackling} (2015)	&   V                     &27& Manual annotation                      & Self-report    &Private                  \\
		\textcircled{17}: CHI-MEI \cite{huang2016unipolar}(2016)	&V &53&DSSS, HAM-D&Clinical
		Assessment&Private\\		
		\textcircled{18}: Pittsburgh \cite{dibekliouglu2018dynamic} (2018)	&    A+V                    &57&  DSM-IV, $HAMD\textgreater15 $                      & Clinical
		Assessment        &Public             \\
		\textcircled{19}: BD \cite{cciftcci2018turkish} (2018) &    A+V                    &46&  DSM-V                       & Clinical
		Assessment        &Public             \\
		
		\textcircled{20}: MODMA \cite{cai2020modma} (2020) &    A+EEG                    &{\tabincell{c}{EEG-128(53), \\ EEG-3(55), \\ A(55)}}&  PHQ-9                       & Clinical
		Assessment        &Public             \\		

		\hline               
	\end{tabular}
	
	\label{table:database_depression}
\end{sidewaystable*}

\section{Depression Databases}\label{sec:Databases}

\begin{figure}[h]
	\centering
	\includegraphics[scale=0.5]{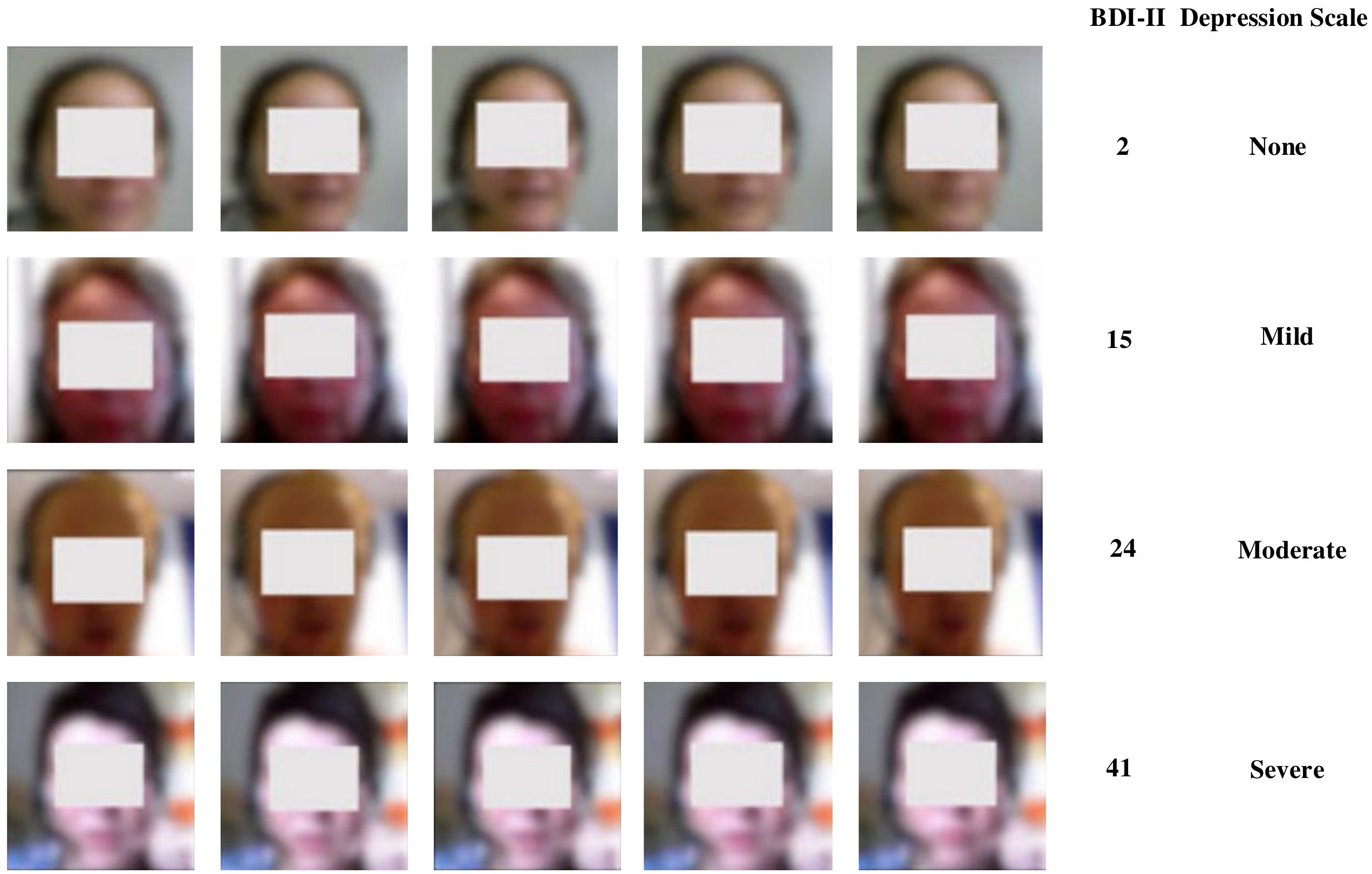}
	\caption{Randomly selected images from video clips along with their \ac{BDI}-II and depression severity scores on the AVEC2014 database \cite{valstar2014avec}. To protect the privacy of participants, the images were blurred, and the regions of the eye were occluded. From the different rows, one can see that the severity of depression is enhanced from none to severe in the images.}
	\label{fig:example_face}
\end{figure}

Deep depression recognition requires sufficient data for the training of a discriminative model. Due to sensitivity of depression, data collection is challenging. Accordingly, different research groups have attempted to collect their own database to study assessment tools for depression estimation. Here, we introduce existing databases that have been widely adopted in the reviewed works for depression detection. Moreover, we also cover other privately released databases. Table \ref{table:database_depression} summarizes the mentioned databases, comprising the number of subjects, the annotation scores, ground truth, availability, additional details. Fig. \ref{fig:example_face} shows some example images from the AVEC2014 database. 

\subsection{The Principles of Data Collection}\label{subsec:Principles}

Collecting depression data requires recruiting a number of participants from hospitals or psychological clinics, which is the most challenging part of depression research. As reviewed in the existing studies, the depressed subjects or health controls are assessed by complying with DSM-IV \cite{american2013diagnostic} standard \cite{siegle2011remission,silk2007pupillary,wang2014pupillometry,cohn2009detecting,girard2014nonverbal,joshi2013can} and$/$or \ac{HAMD} \cite{cohn2009detecting,girard2013social,joshi2013can,joshi2013relative}. Also, Mini International Neuropsychiatric Interview (MINI) \cite{alghowinem2016multimodal} was adopted to diagnose the severity of depression, and QIDS-SR was adopted for defining it. \ac{BDI} has been applied extensively for predicting the scale of depression \cite{siegle2011remission}. In other cases, several standards, PHQ-9 \cite{stratou2013automatic,ghosh2014multimodal,stratou2015automatic,scherer2013automatic,scherer2013audiovisual,gratch2014distress} and \ac{BDI}-II \cite{valstar2013avec,valstar2014avec} are aimed to assess the symptoms related to depression. Furthermore, other recruitment approach (e.g., flyers, posters, social networks, personal contacts, and mailing lists) have been utilized in several studies.

In order to obtain valuable patterns for depression prediction, the experimental environment should be finely designed. In general cases, some agreements were signed prior to the experiment. If the data collection occurs in hospitals, then some devices should be arranged first (e.g., cameras, microphones, and sensors), and the details of variable to record should be planned. Next, the details of the participants should be gathered  (e.g., anamnesis record sheet and cognitive abilities). For instance, in \cite{moore2003analysis}, patients had to satisfy the following standards: 1) diagnosis with MDD or other mental disorder; 2) ability to understand and satisfy the protocol requirements; 3) no other clinical background that could disturb the results (e.g., delirium, dementia, amnestic or other symptoms); 4) no bipolar disorder symptoms assessed; 5) not the criteria of DSM-IV satisfied in the past three months; 6) ability to understand American English. In addition the the healthy control group had to comply with the same principles. Specifically, the healthy individuals should not have had any symptoms related to depression in the past year. All audio samples were collected in the same year and in an identical environment (i.e., room and other experimental setups).    

As mentioned above, the collection environment or setup is essential for recording the data. In some scenarios, emotion elicitation was adopted to to produce a particular emotional response in the participants, which is different between healthy controls and depressed subjects \cite{cummins2015review}.  Moreover, interviews have also been used to discover symptoms of depression, and some spontaneous emotional patterns closely related to depression \cite{mcintyre2009approach} have also been identified during the interviews. Overall, interviews have been conducted by clinicians, psychologists, psychiatrists, and virtual human interviewers and also piloted by a computer to generate several data samples.  

Regarding modality, speech and video samples \cite{cohn2009detecting,ghosh2014multimodal,williamson2014vocal,scherer2013automatic,yu2013multimodal,morency2015simsensei,scherer2013audiovisual,gratch2014distress,scherer2014automatic,gupta2014multimodal,joshi2012depression,joshi2013automated,joshi2013can,meng2013depression,cummins2013diagnosis,joshi2013multimodal,ooi2014early,sidorov2014emotion,kachele2014inferring}, physiological signals \cite{gratch2014distress,meftah2012detecting,meftah2012detecting,acharya2018automated,zandvakili2019use,kan2015decrease,zhang2019multimodal,cai2018case,zhi2018abnormal}, and text \cite{gratch2014distress,cohn2010social} have been employed to improve the performance of depression assessment. However, different modalities were determined by the device used in the data collection stage. For an audio clip, a computer or laptop has been used to record the data samples (i.e., AVEC2013 \cite{valstar2013avec}, AVEC2014 \cite{valstar2014avec}, AVEC2016 \cite{valstar2016avec}). For the video modality, the number of cameras and other attributes (i.e., colors and angles, etc.) have been used; for instance, the face and the whole body have been recorded separately with multiple cameras from different angles \cite{cohn2009detecting}. Also, thermal images based on eye temperature have been adopted to predict the severity of depression \cite{maller2016using}. Microsoft Kinect has also been used to record for the upper body of a participant \cite{stratou2013automatic,scherer2013audiovisual}. The distance between participants and is approximately one meter. A portable three-electrode EEG device has been designed to collect the electroencephalography data \cite{cai2018case}. Similarly, identical to emotion elicitation, the detailed setup alters across different works.

\subsection{Reviewed Databases}\label{subsec:databases}

1 -- DementiaBank database \cite{becker1994natural} contains 226 persons from 2 longitudinal clinical-pathological studies in 4 years in USA between 1994 and 1998, and can be shared for the purpose of research. Video samples were recorded in this database. In the DementiaBank, all persons were diagnosed to suffer Alzheimer's Disease (AD) at some stage. Besides, \ac{HAMD} depression scores were labeled for the part of the participants from the subset of the database.

2 -- The database \cite{stassen1998speech} contains 43 depressed subjects (23 females and 20 males) from the hospital in 1998. Speech samples were recorded before the psychiatric exploration. This database is only accessed by themselves. In addition, most of the patients (79.1\%) represented the onset of improvement in the 12 days, and they had nothing to do with the background of depression severity.   

3 -- The database \cite{france2000acoustical} comprises of 115 peoples in USA. Audio modality is collected from different scenarios. This database is only accessed for lab usage by themselves. The male consisted of seventeen dysthymic patients, twenty-one major depressed patients, as well as ten control subjects. Of these participants, twenty-two high-risk suicidal patients and twenty-four control subjects. F0 and formants features were extracted from audio samples. 

4 -- The database \cite{alpert2001reflections} consists of a 12-week double-blind treatment trial in USA. Speech features were extracted in this database. 10 depressed subjects and 19 health controls were recruited. Depressed subjects were segmented into two groups by complying with the ratings, i.e., retarded or agitated. This database is also for laboratory use only. 

5 -- This database \cite{moore2004comparing} was collected from at the Medical College of Georgia (MCG) at Augusta University in United States. It comprises 33 subjects in this task. Audio samples are collected in this database. All subjects should speak American English fluently and have not abused a significant substance. For each subject, their speech was recorded when reading a short story. Each speech recording fell to 65 separate sentences over the entire session. This database is also for laboratory use only. 

6 -- In this database \cite{yingthawornsuk2006objective}, three different patient groups (i.e., depressed subjects, high–risk suicidal patients, and remitted patients) were included. Speech data samples were recorded of every subject. In addition, Power Spectral Densities (PSD's) features were extracted from the speech data samples. This database is also for laboratory use only. 

7 -- This database \cite{cohn2009detecting} contains 57 participants (20 males and 37 females)  from a clinical trial to diagnose depression symptoms in USA. All of them should met DSM-IV criteria of MDD during the clinical interview step. The severity symptom of depression is assessed on four occasions at approximately seven weeks intervals by a clinical interviewer. HRSD \cite{hamilton1960arating} is adopted to restrict the standard of interviews. Audio and video samples were recorded of the participants in this task.

8 -- ORI database \cite{maddage2009video} comprises of eight participants during family interaction activities. Three interactions (Problem-solving Interaction (PSI), Event Planning Interaction (EPI), and Family Consensus Interaction (FCI)) were performed in the sessions. Every adolescent was asked to record a one-hour video recording during the sessions. The age of adolescents ranged from 12 to 19. The eight subjects fell into two groups, i.e., depressed subjects (4) and non-depressed groups (4). Of all of the participants were selected with white, and nobody wore glasses in the sessions. This database is not opened for public use.

9 -- ORYGEN database \cite{ooi2011prediction} collects from ORYGEN Youth Health Research Center in Australia. The video and audio data samples are recorded from the discussions between parents and their children (the age ranging from 12 to 13 years) by two different family interactions (i.e., EPI, PSI). The mentioned video samples were recorded in two steps. In the first step (T1), 191 non-depressed subjects (94 females and 97 males) were recruited. Two years later, 15 (6 males \& 9 females) of 191 suffered from MDD and three participants (1 male \& 2 females) suffered from Other Mood Disorders (OMD) in the second step (T2). This database is not opened for public use. 

10 -- BlackDog database \cite{Alghowinem2012From} is collected from an organization named BlackDog institute, focusing on a clinical study in Sydney, Australia. 80 participants (the age ranging from 21 to 75) participated. To ensure the availability of experiments, all participants had to comply with the criteria of DSM-IV. Speech data are recorded during the conversation between the interviewer and participants. The clinical interaction was performed by asking specific questions (eight groups), in which participants were required to describe events stimulated by specific emotions.

11 -- This database \cite{mundt2012vocal} comprises 165 adults (104 females and 61 males) with major depression disorders between November 2006 and August 2007 in the united states. Speech samples were recorded with automated telephone devices in this database. All participants should meet the following criteria: 1) the age between 18 and 65; 2) not taking psychotropic
medications; 3) \ac{HAMD} of 22 or greater; 4) Symptoms of MDD has been diagnosis with DSM-IV, and have been lasted for a month. This database is also used in their own study.

12 -- AVEC2013 database \cite{valstar2013avec} refers to a selection from the audiovisual depressive corpus, covering $340$ videos from $292$ peoples by performing a human-computer interaction. $31.5$ years ($18$ and $63$ years) was the average age of the participants. \ac{BDI}-II was adopted for labeling every audio and video segment. In this database, the organizer only provided a total of 150  audio and video clips, falling to three equivalent partitions ( training, development, and test set). Different from the above mentioned databases, AVEC2013 is open for researchers to design the ADE systems.  

13 -- AVEC2014 corpus~\cite{valstar2014avec} was divided from the AVEC2013. The only difference was that the AVEC2014 database contained two tasks, i.e., Freeform and Northwind. Thus, every partition covered 100 data samples, respectively. Therefore, AVEC2014 contains 300 data samples in total. \ac{BDI}-II was used for labeling every audio and video clip. 

14 -- Crisis Text Line. A web python development tool, back-end Flask, was adopted to store the data using MySQL database \cite{chen2014visualizations}. A single URL of the website was exploited to call with Flask routing between the front-end and back-end through AJAX. The interaction occurred between Flask and MySQL database using a Python class. Furthermore, the Python class could process the generated data accessing into and out of the database. Text features are extracted in this database.

15 -- DAIC database \cite{gratch2014distress} was collected based on a semi-structured clinical interactions in USA. Four types of interviews were conducted, i.e., Face-to-Face, Teleconference, Wizard-of-Oz, and Automated. The database contains 189 sessions of interactions , and consists of the audiovisual cues, as well as physiological data (e.g., galvanic skin response (GSR), electrocardiogram (ECG), and respiration). In addition, text modality was also collected during the interactions. Different verbal and non-verbal features were used to annotate the corpus. DAIC is the same as AVEC2013 and AVEC2014 to open access for the researchers.   

16 -- Rochester database \cite{zhou2015tackling} consists of 32 participants under differ conditions. The normal group involved 27 participants (16 females and 11 males) in this database. The age of this group ranged from 19 to 33. Also, 5 depressed patients (2 severe and 3 moderate) and a healthy control group of five participants were covered.

17 -- In the CHI-MEI database \cite{huang2016unipolar}, six discrete videos (i.e., disgust, fear, sadness, surprise, anger, and happiness) were adopted to arouse the subjects to express their expressions based on their facial region and the responses speech of them. The CHI-MEI speech database was collected from the speech response of the subjects by a clinician in
CHI-MEI Medical Center, Taiwan. The audio and video data were collected in this dataset. In total, 15 BDs, 15 UDs and 15 healthy controls are recruited in CHI-MEI. In addition, the participants had to complete a baseline recording before data collection. Afterward, the participants watched six emotional videos.

18 -- Pittsburgh database \cite{dibekliouglu2018dynamic} comprises of 57 (34 females, 23 males) depressed participants from a clinical treatment for depression. The age ranged from 19 to 65 years (mean=39.65). All the participants had to reach DSM-IV criteria for MDD. Severity of MDD was assessed at 1, 7, 13, and 21 weeks by 10 random clinical interviewers. This database is also open for public usage. 

19 -- BD database \cite{cciftcci2018turkish} consists of 46 patients and 49 healthy controls of the mental health service of a hospital. To gather the sociodemographic and clinical patterns, all patients should perform semi-structured interviews by the SKIP-TURK. Young Mania Rating Scale (YMRS) and MADRS were employed to estimate the depressive and manic features in the next following days (0, 3, 7, 14, 28), and then altered in the third month. During this step, audiovisual samples were recorded. Accordingly, every video session was annotated by bipolar mania/depression ratings. This database is used as the chanllenge data in AVEC2018. 

20 -- MODMA database \cite{cai2020modma} was collected from audio and EEG signals for mental disorder analysis in China. Experienced psychiatrists rigorously recruited all the participants from hospitals. The EEG database contains data samples recorded with traditional 128-electrodes mounted elastic cap as well as a new wearable 3-electrode EEG recorder for pervasive usage. The 128-electrodes EEG signals were recorded with resting-state and under-stimulation from 53 subjects, while 3-electrode EEG signals were recorded with resting-state from 55 subjects. Specific to the audio data, the samples were collected from 52 subjects by allowing participants to be interviewed, read stories, and watch the emotional pictures. 

According to the mentioned databases, the following discussions are made:

\begin{itemize}
	\item From the perspective of openness, most of the databases were only used for their own research and not released publicly for depression recognition study. Only few databases were released publicly for depression recognition, i.e., AVEC2013 and AVEC2014 \footnote{http://avec2013-db.sspnet.eu/}, DAIC-WOZ \footnote{ http://dcapswoz.ict.usc.edu/}, Pittsburgh dataset \footnote{http://www.pitt.edu/emotion/depression.html}, and MODMA dataset \footnote{http://modma.lzu.edu.cn/data/index/}. The rest of the databases may be available for the researchers in some scenarios.  
	
	\item Most of the databases were collected by the region of the US and EU. There is only one database available for researchers in China, which is MODMA. 
	
	\item From modalities, most of the database involved one or/and more (e.g., audio, video, physiological signals, text). 
	
	\item In terms of the number of subjects, all the databases consisted of limited data samples, which is explained as depression is a mental disorder and also kept as a secret by the depressed subjects. 
	
\end{itemize}

\section{Deep Audiovisual Depression Recognition}\label{sec:methods}

This section presents the common procedures adopted in \ac{ADE}, i.e., preprocessing, deep feature extraction, and classification/regression. For instance, we describe raw audio and other audio features that are used broadly as input for depression recognition, as well as  other data modalities (e.g., video, text from transcripts, and physiological signals). In the following, the works introduced in the literature are separated into three groups: 1) deep \ac{ADE} networks for audio modality; 2) deep \ac{ADE} networks for static images, and 3) deep \ac{ADE} networks for image sequences. In addition, different network types are also introduced for the mentioned groups along with discussion. 

\subsection{Preprocessing}\label{subsec:preprocessing}  

Both conventional and end-to-end schemes require some pre-processing steps before the actual depression recognition and analysis.   

For the audio data, the sample rate is usually processed to 16 kHz or others (e.g., AVEC2013). Meantime, to generate the spectrograms of audio data, Discrete Fourier Transform (DFT) method is adopted for conducting a Time-Frequency (TF) characterization for audio signals. To select DFT parameters, a Hanning window (23ms with 50\% overlap) is used \cite{he2018automated}. In addition, to extract efficient hand-crafted features, the length of low-level descriptor (LLD) is also considered in depression recognition studies. In \cite{he2018automated}, they tried LLDs of different lengths and suggested that 20s is sufficient to obtain a good performance. In \cite{niu2020multimodal}, the authors sampled the waveforms at 8KHZ and generated the 129-dimensional normalized amplitude spectrogram using a short-time Fourier transform with 32 ms Hamming window and 16 ms frame shift for AVEC2013 and AVEC2014 databases.

For video data, image normalization, face detection, and alignment between the adjacent frames, are commonly used preprocessing techniques. Viola and Jones proposed a general algorithm for face detection \cite{viola2004robust}. Besides, the OpenFace toolkit provides a free tool for face detection and alignment in many applications \cite{baltruvsaitis2016openface}. The computer expression recognition toolbox is used in numerous fields, but it is not free to use at the moment \cite{littlewort2011computer}. Moreover, a comprehensive facial pre-processing tool can be found on the web \footnote{http://nordicapis.com/20-emotion-recognition-apis-that-willleave-you-impressed-and-concerned/}. In addition, video data is pre-processed with different types for depression recognition, i.e., frame-level images, and image sub-sequences, and image sequences.

\subsection{Deep ADE Networks for Audio Modality}\label{sec:Audio Modality}

In the databases mentioned above, the extraction of hand-crafted features stays dominant in audio-based \ac{ADE}. Next we describe hand-crafted feature extraction from audio cues for \ac{ADE}. 

Since 1998, a range of feature representation methods have been proposed to estimate the severity of depression. Here we list only some classical (shallow) methods for depression recognition, and then focus primarily on the deep automatic depression recognition methods. In 1998, the pause duration of speech was tightly related to the \ac{HAMD} scores for 60\% of patients \cite{stassen1998speech}. In 2001, Alpert et al. \cite{alpert2001reflections} found that there are differences between healthy controls and depressed individuals. Cannizzaro et al. found an important relationship between reduced speak rate and \ac{HAMD} score in 2004 \cite{cannizzaro2004voice}. In addition, they found that different acoustic features also could impact the representation of depression (e.g., percent pause time, speaking rate, and pitch variation). It is noteworthy that variations of speaking rate and pitch were considered an important representation for depression analysis. In 2008, Moore et al. \cite{moore2007critical} studied the combination of a wide range of features, e.g., prosodic, voice quality, spectral, and glottal. They obtained comparable performance for classifying the absence/presence of depression \cite{moore2007critical}. Many LLD indicators (e.g., prosodic, source, formant, and spectral) have been identified as efficient predictors of depression. For an in-depth review of speech-based depression recognition, please refer to the article \cite{cummins2015review}. As revealed from the review work, one can note that hand-crafted features have achieved promising performance for depression prediction. However, some issues remain; for instance, manual work and expert knowledge are significant for feature selection, which wastes labor resources. Furthermore, representations learned via \ac{DL} have exhibited excellent performance compared to hand-crafted in multiple disciplines, and \ac{ADE} is not an exception.

\begin{figure}[h]
	\centering
	\includegraphics[scale=0.8]{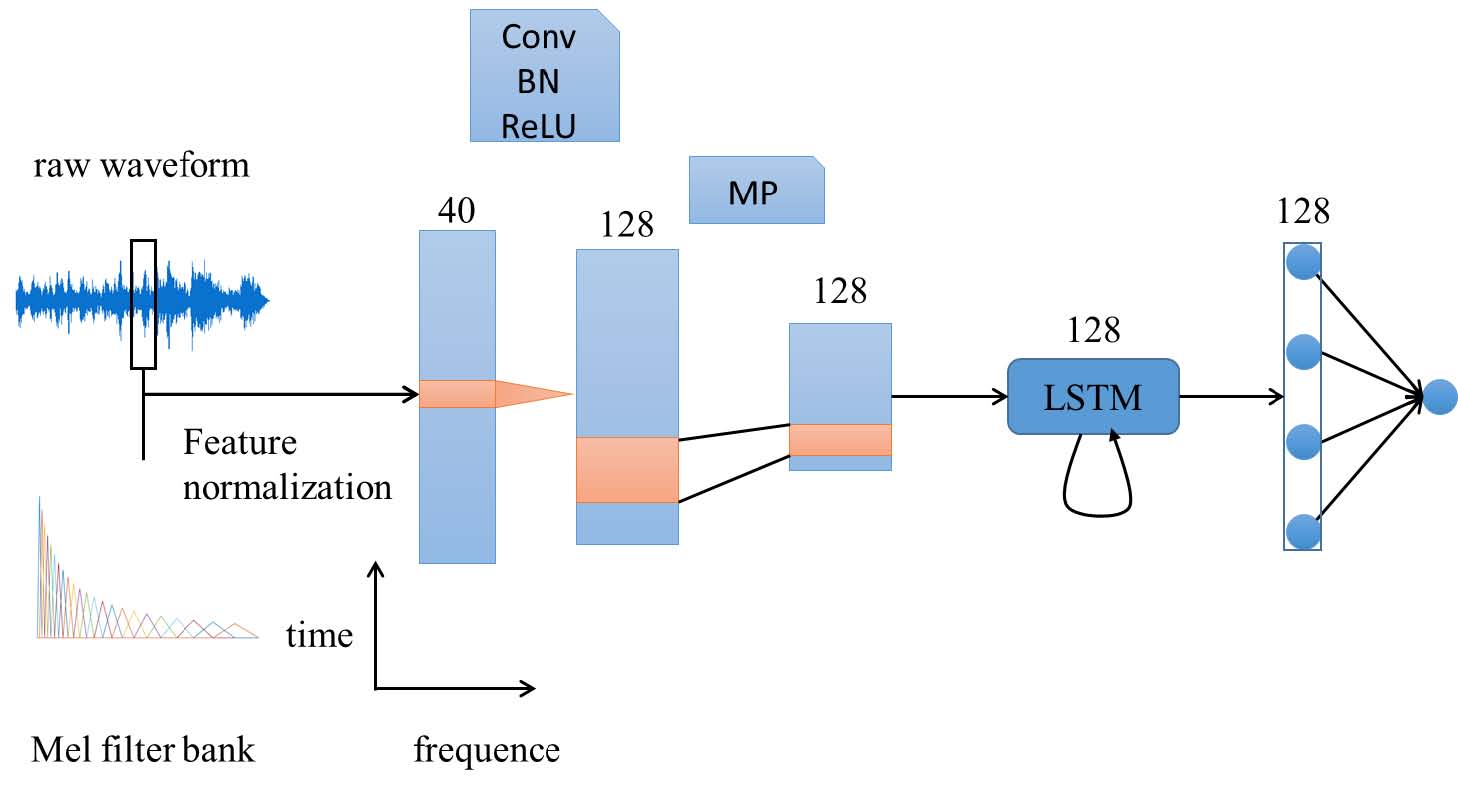}
	\caption{The pipeline of DepAudioNet framework from \cite{ma2016depaudionet}. 
In the framework, CNN and LSTM are combined to model different scale features for audio-based depression recognition. The DCNN can model the high-level patterns of the raw waveforms. The LSTM can learn the combination of short-term and long-term representations from Mel-scale filter bank features. Mel-scale filter bank feature is used as a LLD to represent the characteristics from the vocals. Conv is the convolutional operation, BN represents the batch normalization, ReLU is the rectified linear unit operation, and MP is the multi-layer perceptron.}
	\label{fig:DepAudioNet}
\end{figure}

\begin{figure}[h]
	\centering
	\includegraphics[scale=0.6]{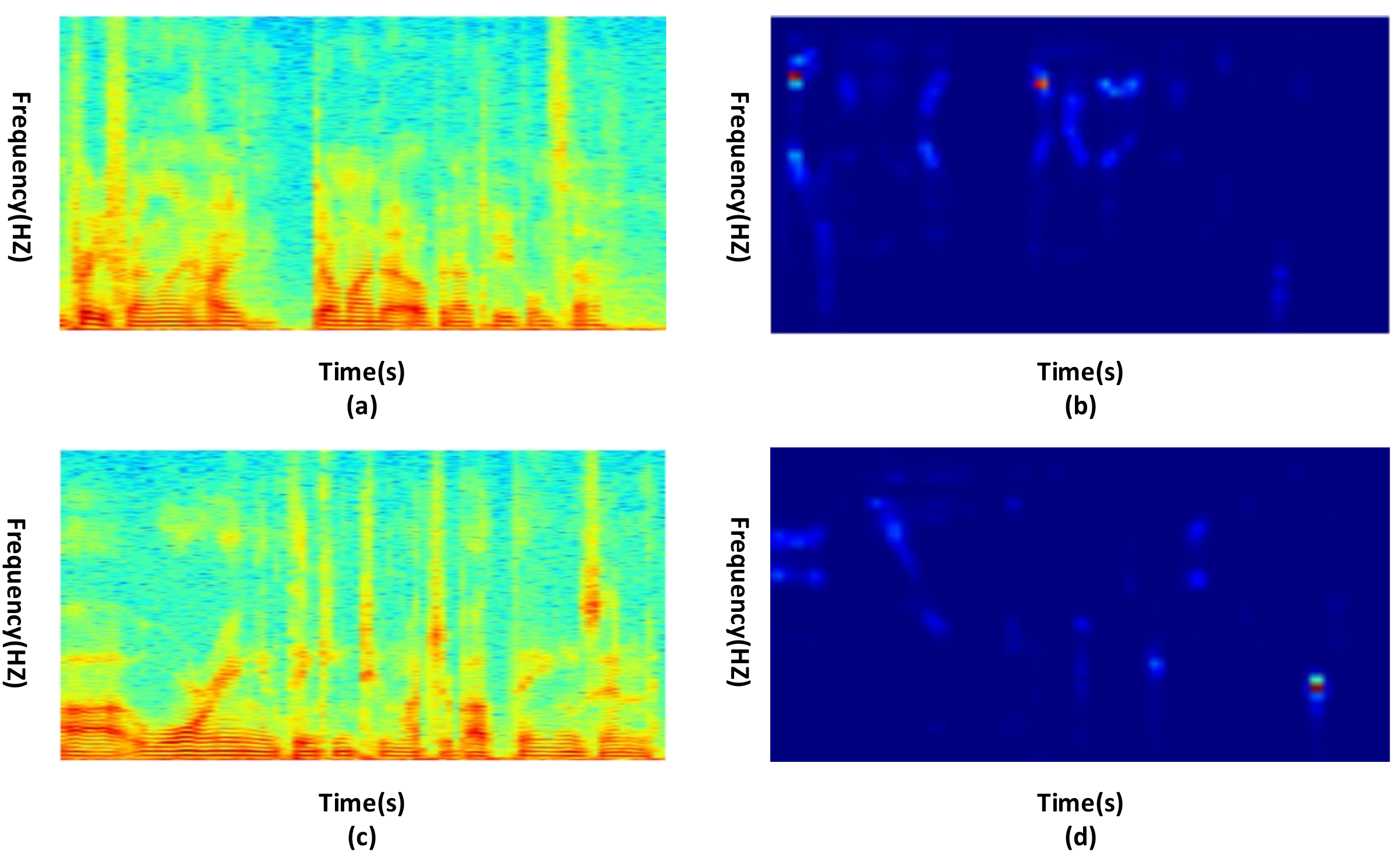}	\caption{The visualization of spectrogram and mel-scale filter banks. (a) and (b) represent the spectrogram and filter bank features from an audio segment of a health control. (c) and (d) show the spectrogram and filter bank features from an audio segment of a depressed individual \cite{ma2016depaudionet}.}
	\label{fig:DepAudioNet_spectrogram}
\end{figure}

\begin{figure}[h]
	\centering
	\centerline{\includegraphics[scale=0.85]{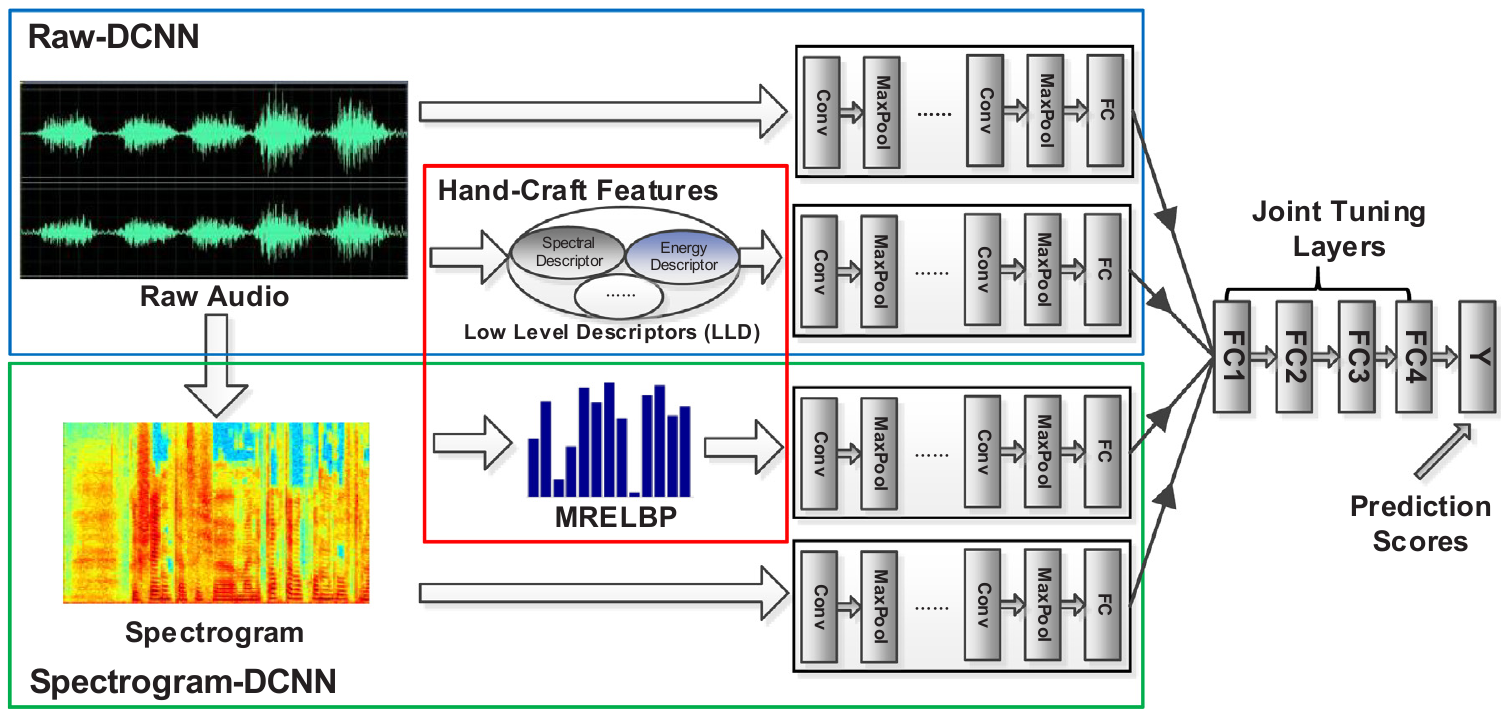}}
	\caption{The framework for depression proposed in \cite{he2018automated}. The four-stream deep features, i.e., hand-crafted features (LLD, MRELBP), deep-learned features (raw audio, spectrogram), are fused for deep depression recognition. Raw-\ac{DCNN} (Top) adopts LLD and raw audio signals as input, whilst Spectrogram-DCNN (Bottom) utilizes MRELBP and spectrogram features as input. The red box represents Hand-Crafted features. The other two arrows denote deep-learned features. The \ac{BDI}-II score is calculated by averaging and aggregating the output of the four \ac{DCNN} branches. } 	
	\label{fig:2_pipeline}	
\end{figure}

In 2016, \cite{ma2016depaudionet} proposed a new model based on deep learning, i.e., DepAudioNet, to mine the depression representation from vocal cues, adopting LSTM and \ac{DCNN} to encode a discriminative audio representation for depression recognition (see Fig. \ref{fig:DepAudioNet}). \ac{DCNN} can model the spatial feature representations from the raw waveforms, and LSTM can learn the short-term and long-term feature representation from the mel-scale filter banks \cite{shannon2003comparative}. In addition, to balance the positive and negative samples, a random sampling approach is adopted in the model training stage before using LSTM. Using the DepAudioNet, different scale representations, i.e., high-level, short-term, and long-term features, are extracted. To further explain different representations between healthy controls and depressed subjects, Fig. \ref{fig:DepAudioNet_spectrogram} provides a comparison of the spectrogram and filter bank features extracted from an audio segment. The goal of the authors is to try to use deep learning methods to estimate the severity of depression. Most importantly, despite the small size of the training data, the deep learning methods can also learn the discriminative patterns from audio signals. 

Even though the used depression databases have only a limited number of samples, deep learning-based depression recognition approaches aroused great attention from numerous researchers. Then in 2018, a fusion of deep-learned and hand-crafted features was used, capable of effectively measuring the severity of depression from speech. In the framework, \ac{DCNN} was used to learn and fuse the shallow and deep patterns for evaluating the severity of depression. Specifically, LLD features are extracted by OpenSMILE toolkit \cite{eyben2015opensmile} as hand-crafted features from audio. Median robust extended local binary patterns (MRELBP) are extracted as hand-crafted features from spectrograms. Raw audio and spectrograms are used as input into the \ac{DCNN} to obtain the deep learned features. To learn the complementary representations between the hand-crafted features and the deep-learned features from the Raw-\ac{DCNN} and Spectrogram-\ac{DCNN}, a joint fine-tuning technology is used. In addition, to overcome the issue of the small number of samples, a data augmentation approach is introduced. Most importantly, the proposed scheme presents an end-to-end architecture for depression recognition \cite{he2018automated} (see Fig. \ref{fig:2_pipeline}). The contribution of the work \cite{he2018automated} is the attempt to fuse hand-crafted and deep learned features from speech for depression estimation. Also, the authors \cite{he2018automated} extracted texture features from spectrograms for predicting the severity of depression. The authors validated the proposed method and obtained a good performance on AVEC2013 and AVEC2014 databases with RSME of 10.00 and 9.98, respectively (see Table \ref{table:static images}).

\begin{figure}[h]
	\centering
	\centerline{\includegraphics[scale=0.8]{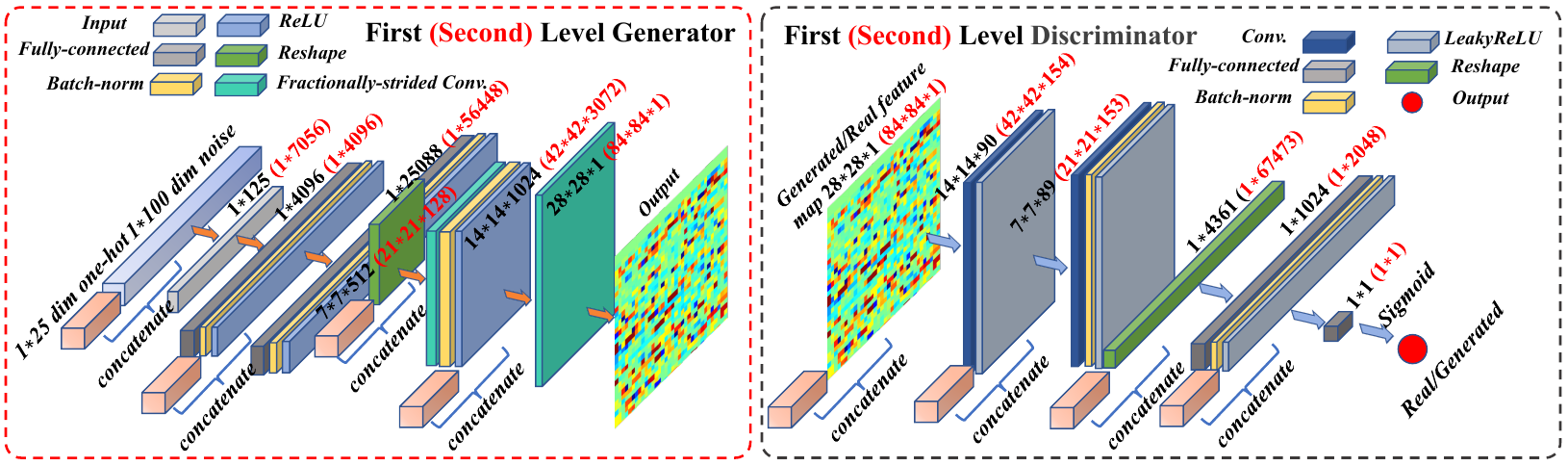}}
	\caption{An illustration of the framework proposed in \cite{yang2020feature}. In the DCGAN framework, a 2-level hierarchical learning strategy is proposed to streamline the training procedure \cite{yang2020feature}. In the first level, feature maps are split into 9 blocks with size $28\times28$. For every block, the model generates synthetic representations. After that, 9 DCGANs are generated with the same architecture. The output of the first level (with the size of  $9\times28\times 28$) is fed into the second level to obtain the global features. }
	\label{fig:yang}	
\end{figure}

Due to the limited sizes of databases available for depression recognition, different studies have proposed augmenting the data somehow. For instance, in \cite{yang2020feature}, Deep Convolutional Generative Adversarial Network (DCGAN) was proposed to augment the size of data samples to enhance the accuracy of ADE task from audio signals. To validate the performance of the augmented features, three measurement criteria have been proposed, i.e., spatial, frequency, and representation learning. The proposed architecture was capable of achieving comparable performance with most of the methods on the DAIC database, with RMSE of 5.52 and MAE of 4.63 (see Table \ref{table:static images}). As illustrated in Fig. \ref{fig:yang}, the DCGAN framework contains a learning strategy with two levels, to improve the convergence speed of the training. In the first level, feature maps were split into 9 blocks with a size of $28\times28$. For every block, a DCGAN model is used to represent synthetic representations. After that, 9 DCGANs are generated with the same architecture. The output of the first level (with the size of  $9\times28\times$) is fed into the second level to obtain the global features. The advantage of this architecture is that complex training is transformed to a more straightforward procedure.

Table \ref{table:static images} lists the reviewed approaches for depression recognition with audio cues. Table \ref{table:static images} shows that many studies adopted the \ac{DCNN} model to extract deep features for predicting the depression scale. From the algorithmic perspective, researchers adopted the widely used \ac{DL} technologies for evaluating the severity of depression, i.e., \ac{DCNN}, LSTM, etc. It is noteworthy that the raw audio signals are fed into the \ac{DCNN} directly to overcome the drawbacks of the conventional feature design methods \cite{he2018automated}. Furthermore, Niu et al. \cite{niu2020multimodal} attempted to convert the audio segments into a spectrogram to feed into the deep architecture. They sampled the audio clips at 8KHZ, and adopted STFT using a Hamming window with 32ms and shift with 16ms to generate 129-dimensional spectrograms on the two AVEC2013 and AVEC2014 databases. In addition, they found that the optimal spectrogram length is 64 frames (1s) with the shift of 32 frames (0.5s) of the two databases, respectively. 

In 2021, Niu et al. \cite{NIU2021} proposed a novel framework that integrates with the Squeeze-and-Excitation (SE) component and a Time-Frequency Channel Attention (TFCA) block to represents informative characteristic related to depression. Additionally, to consider the time-frequency features of the data, a Time-Frequency Channel Vectorization (TFCV) block is proposed to form the tensor. Moreover, they integrated the introduced blocks (i.e., TFCA and TFCV blocks) and the two blocks (i.e., Dense block and Transition Layer) of the DenseNet into a unified framework to generate Time-Frequency Channel Attention and Vectorization (TFCAV) network. The contribution of this work \cite{NIU2021} is that time-frequency attributes are considered to learn the informative patterns from spectrograms. The performances of the introduced method obtained on AVEC2013 and AVEC2014 with RMSE of 8.32 and 9.25, respectively.  

\begin{figure}[h]
	\centering
	\centerline{\includegraphics[scale=1]{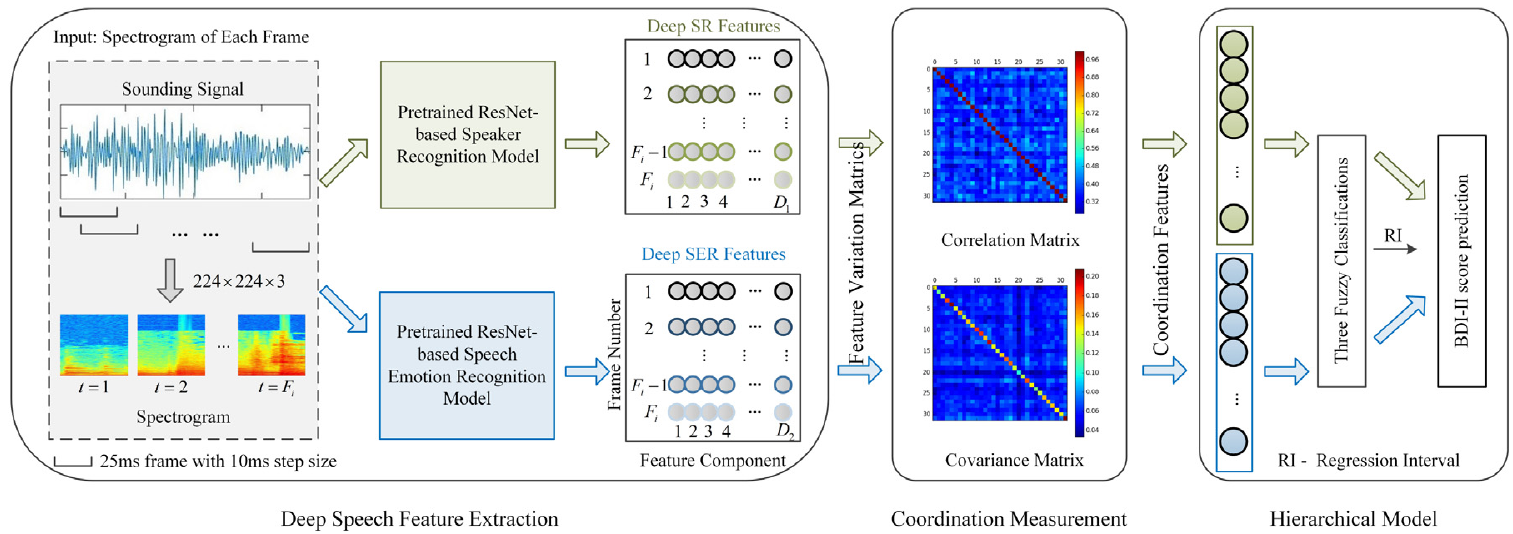}}
	\caption{Illustration of the architecture for \ac{ADE} from speech proposed in \cite{dong2021hierarchical}. The architecture can be divided into three steps: deep speech feature extraction, coordination measurement, and hierarchical model construction. In the first step, a spectrogram is used for learning the features of the frame-level speaker recognition (SR) and speaker emotion recognition (SER)  from the pre-trained SR and SER models. In the second step, the FVCM algorithm is adopted to compute the correlation and covariance coefficients of the time-delayed multi-channel variations to obtain the coordination features. In the third step, a ADE model is introduced.}
	\label{fig:SR_SER}	
\end{figure}

In \cite{dong2021hierarchical}, the authors proposed a deep architecture for \ac{ADE} from speech with two contributions. The first one is that Speaker Recognition (SR) and Speaker Emotion Recognition (SER) features were fused to improve \ac{ADE}'s performance. The second contribution is that the Feature Variation Coordination Measurement (FVCM) algorithm is used to model the correlation and covariance coefficients of the time-delayed multi-channel variations (see Fig. \ref{fig:SR_SER}).  
\begin{figure}[h]
	\centering
	\centerline{\includegraphics[scale=0.4]{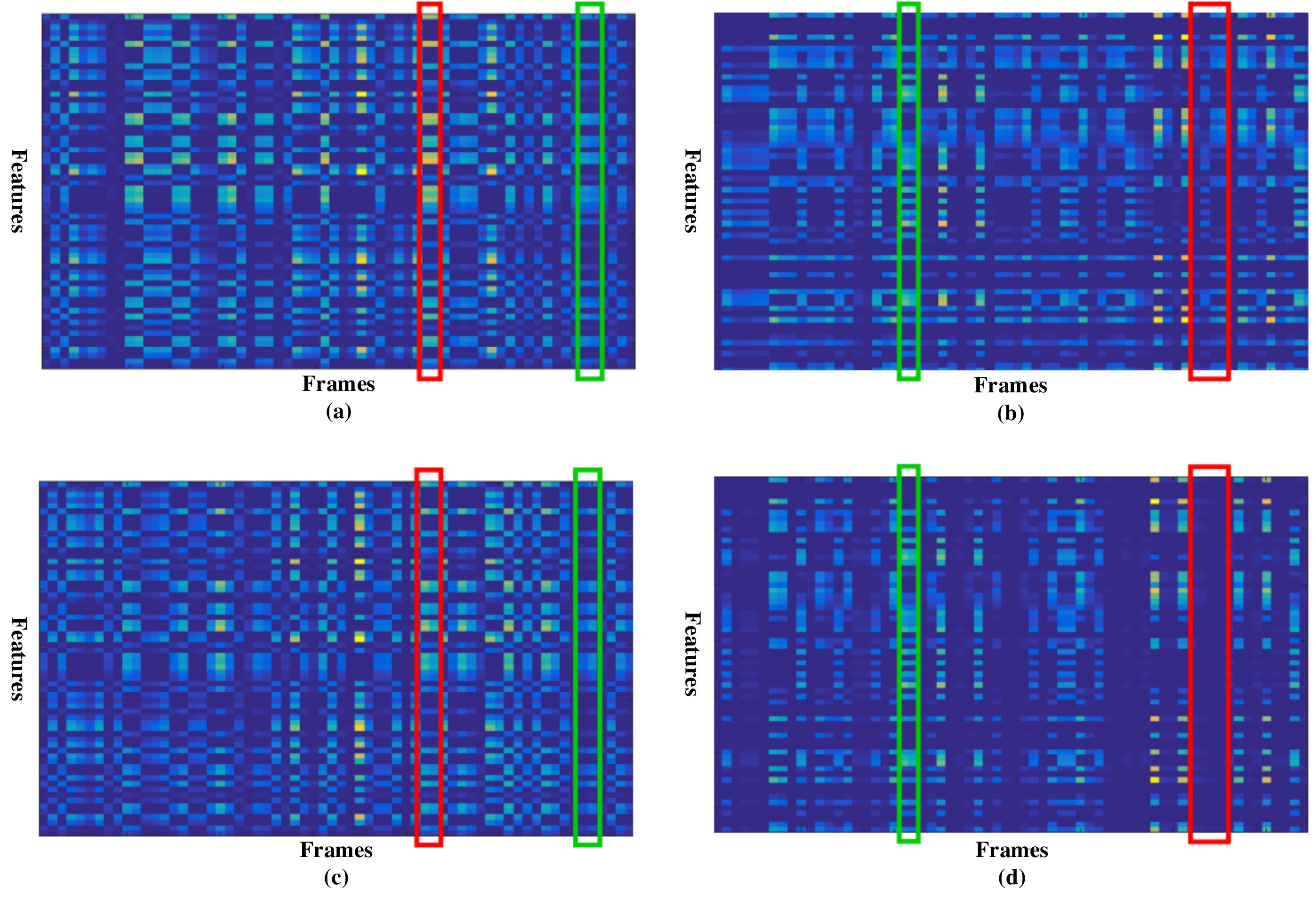}}
	\caption{The example of audio and video features for health controls and depressed subjects. (a) and (b) represent the audio and video characteristic for the healthy controls (No. 203-1) of the AVEC2013 database, respectively. (c) and (d) represent the patterns for the depressed subjects (No. 236-1) of the AVEC2013 database. The scale of depression of No. 203-1 is none (3), while No. 236-1 is 23 (moderate). To further illustrate the depression scales, red and green rectangles are used to represent discriminative and less discriminative feature vectors. }
	\label{fig:audio_video_visulize}	
\end{figure}

Fig. \ref{fig:audio_video_visulize},  (a) and (b) illustrate the audio and video sub-sequences for a healthy control.  (c) and (d) represents  the same patterns for a depressed individual. Imagesc toolkit of MATLAB is adopted to plot the figures. From the Fig. 10, one can see that there are difference between (a) and (c) or between (b) and (d). For instance, the blocks enclosed by red rectangles are discriminative between the individuals, whereas the parts covered by green rectangles more are similar. Based on Fig. \ref{fig:audio_video_visulize}, we can make the following observations: The discriminative patterns of audio and video frames have different contributions for healthy controls and depressed subjects. The mentioned studies provide motivation for the subsequent works in ADE.

\subsection{Deep ADE Networks for Video Modality}\label{sec:Video Modality}

Besides the audio modality, visual cues also important for deep depression recognition. Therefore, various researchers from the affective computing field have explored the discriminative patterns in videos for \ac{ADE}. In the following, we introduce the studies based on video for \ac{ADE}. Accordingly, we divide the mentioned methods into two groups based on the input of the deep networks: deep \ac{ADE} networks for single image and deep \ac{ADE} networks for image sequences.   

\subsubsection{Deep ADE Networks for Single Image}\label{sec:Video 2D}

\begin{figure}[h]
	\centering
	\centerline{\includegraphics[scale=0.75]{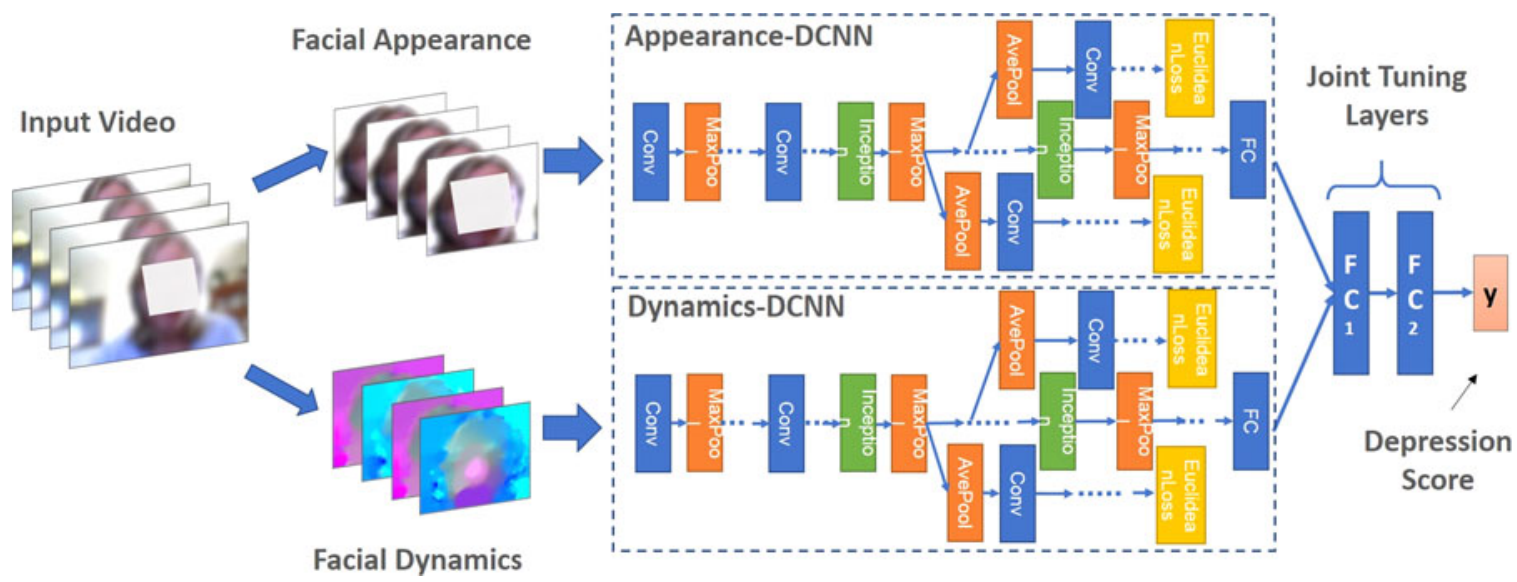}}
	\caption{Illustration of the model to predict the depression state using deep networks from \cite{zhu2017automated}. For the first branch, facial images are fed into the appearance \ac{DCNN} to get static feature representations. In the second branch, optical flows are input into the dynamics \ac{DCNN} to model facial dynamics. Then the final \ac{BDI}-II score is generated via pooling (i.e., averaging and aggregating) the two outputs of each frame from both branches.}
	\label{fig:Zhuyu}	
\end{figure}

\begin{figure}[h]
	\centering
	\centerline{\includegraphics[scale=0.8]{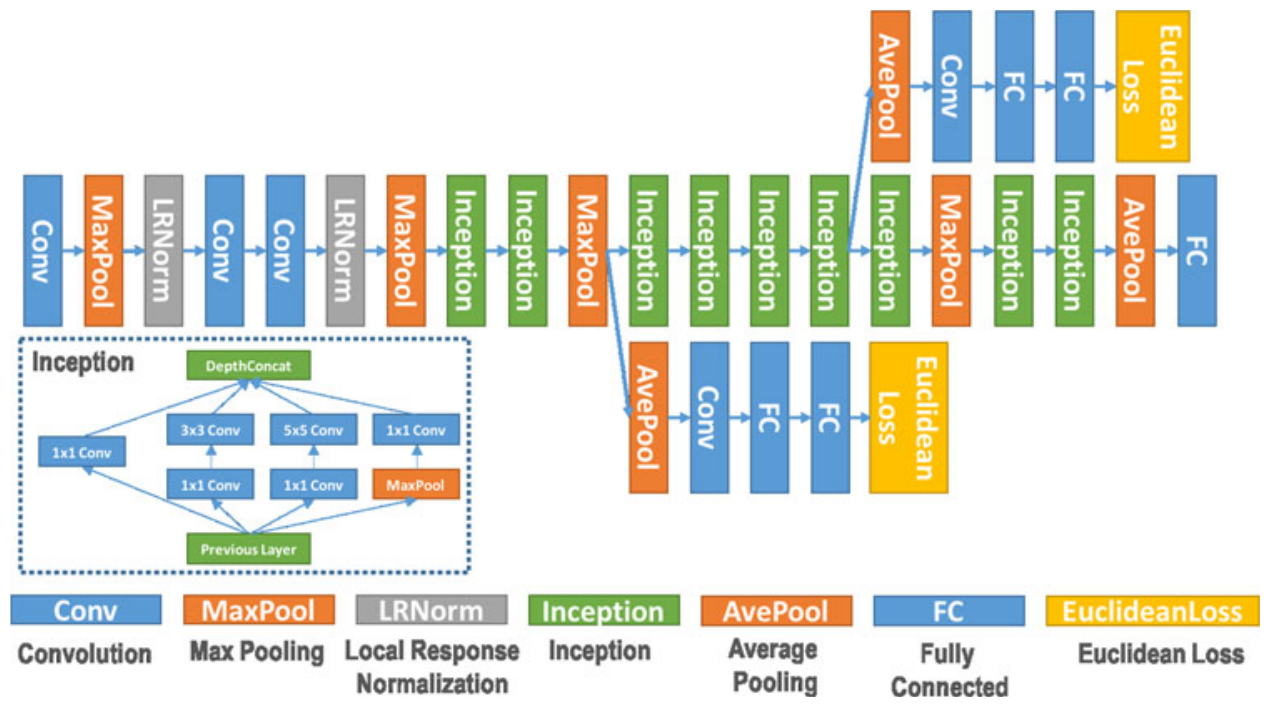}}
	\caption{The detailed architecture of Appearance-\ac{DCNN} and Dynamics-\ac{DCNN} of Fig. \ref{fig:Zhuyu} for ADE. Different pre-processing and pre-training methods are used for the above two branches, and the same deep network architecture is used to learn the features. After that, the predictions of the two architectures are fused to obtain the ensemble \ac{BDI}-II scores.}
	\label{fig:Zhuyu2}	
\end{figure}

The study \cite{zhu2017automated} was an initial attempt to adopt deep learning for depression detection from static images. A two-stream network was developed in their proposed framework to use facial images and optical flow features to learn the depression patterns (Fig. \ref{fig:Zhuyu}). Appearance-\ac{DCNN} and Dynamics-DCNN have been introduced to model the static and dynamic patterns for depression recognition. The Appearance-DCNN includes two steps. The first step consists of training a model from scratch on a public CASIA WebFace Database with 494,414 images from 10,575 subjects \cite{yi2014learning}. After that, the deep model includes the discriminative representations related to facial structures, which can provide sufficient information for the \ac{ADE} task. However, the pre-trained model can not be directly used for \ac{ADE}. The second step is to fine-tune the pre-trained model for \ac{ADE}. However, the \ac{ADE} task based on AVEC2013 and AVEC2014 can be viewed a regression problem from the perspective of machine learning. Hence, the softmax loss function is changed into the Euclidean loss for \ac{ADE}. To further model the dynamics between several consecutive video frames, optical flow displacements are computed for the Dynamics-DCNN. The subtle dynamic patterns and motions of the face are explored, and redundant information of videos is reduced by using the optical flow. In particular, the study leverages the ability of the existing large models to predict the \ac{BDI}-II scores on small datasets. Most importantly, this work \cite{zhu2017automated} has given a certain inspiration for the following works based on deep learning for depression recognition and analysis. The detailed architecture of Appearance-\ac{DCNN} and Dynamics-\ac{DCNN} of Fig. \ref{fig:Zhuyu} for ADE.

\begin{figure}[h]
	\centering
	\centerline{\includegraphics[scale=1]{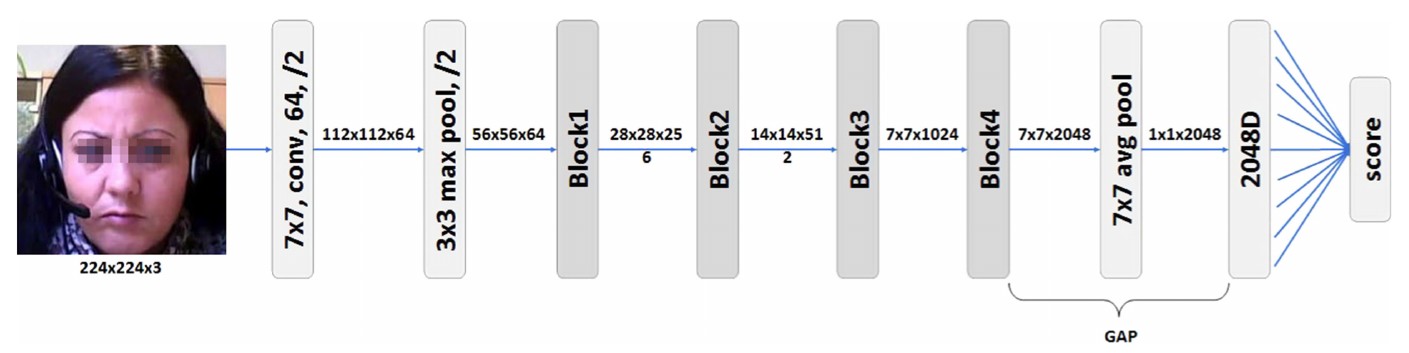}}
	\caption{The detailed architecture of DepressNet for \ac{ADE} task \cite{zhou2018visually}. In this architecture, the facial images are first pre-processed by the OpenFace toolkit to ensure they have the same scale. The architecture has residual connections, similar to those in the popular ResNet architecture. DepressNet contains four blocks, which consist of 3, 4, 6, 3 bottleneck architectures for feature representation. Then the 2048D features are extracted from the architecture for ensemble depression prediction. }
	\label{fig:DepressNet}	
\end{figure}

\begin{figure}[h]
	\centering
	\centerline{\includegraphics[scale=0.9]{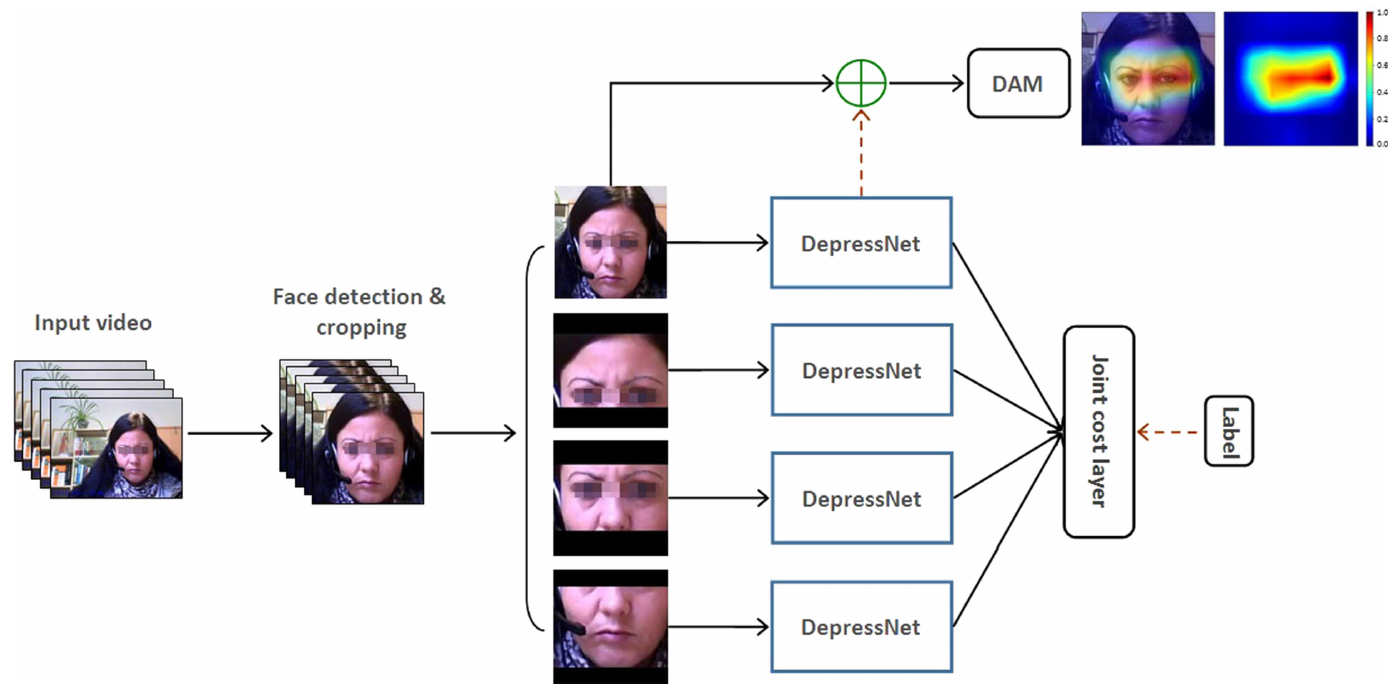}}
	\caption{The detailed architecture of Multi-Region DepressNet for \ac{ADE} task from \cite{zhou2018visually}. In this architecture, the facial images are first pre-processed by OpenFace toolkit to ensure them at the same scale. Then the facial area is divided into different regions, which are fed together with full face into DepressNet to estimate the \ac{BDI}-II score. }
	\label{fig:Multi-DepressNet}	
\end{figure}

In 2018, Zhou et al.\cite{zhou2018visually} proposed a novel deep architecture named DepressNet to learn representations from images for depression recognition, as shown in Fig. \ref{fig:DepressNet}. Different deep architectures (AlexNet, ResNet, GoogleNet) were pre-trained on the CASIA database. DepressNet is constructed by changing the softmax layer into a regression layer, followed by a global average pooling (GAP) layer, as shown in Fig. \ref{fig:DepressNet}. Specifically, the DepressNet consists of four bottleneck blocks, which include 3, 4, 6, and 3 bottleneck structures. After that, the deep model is fine-tuned on the AVEC2013 and AVEC2014 databases. Meanwhile, the loss function also changed into the squared loss, which can be written as follows:
\begin{equation}
L = \frac{1}{{2M}}\sum\limits_{j= 1}^M {{{\left( {{g_{}}({y_j}) - {\ell _j}} \right)}^2}},
\end{equation} 
where $M$ represents the batch size, ${g_{}}({y_j})$ and ${\ell _j}$ are the predicted value and the label of the $j$-th face image of sample $x_j$, respectively.
Another method, a multi-region DepressNet (MR-DepressNet) \cite{zhou2018visually} has been designed for learning different scale models for overall depression recognition, as shown in Fig. \ref{fig:Multi-DepressNet}. In this architecture, to learn the discriminative patterns from different regions and full images, a four-stream DepressNet is proposed. In order to learn more robust representations, the output of the four sub-architectures is combined at the cost function layer. Formally, the loss function of MR-DepressNet can be written as:
\begin{equation}
L = \frac{1}{{2M}}\sum\limits_{j = 1}^M  \left( {\frac{1}{J}\sum\limits_{j = 1}^J {{g_j}({y_j}) - {\ell _j}} } \right)^2
\end{equation}   
Where $M$ represents the batch size, $J$ represent the number of the image regions, ${g_j}({y_j})$ denotes the output of the $j$-th sub-architecture, and ${\ell _j}$ represents the \ac{BDI}-II score of the $j$-th sample ${y_j}$. For the multi-region architecture, the ensemble output of an image was calculated via $\frac{1}{J}\sum\nolimits_{j = 1}^J {{g_j}({y_j})}$. Additionally, as illustrated in Fig. \ref{fig:Multi-DepressNet}, depression activation maps are obtained by learning and fusing the four facial regions. 

\begin{figure*}[h]
	\centering
	\centerline{\includegraphics[scale=0.90]{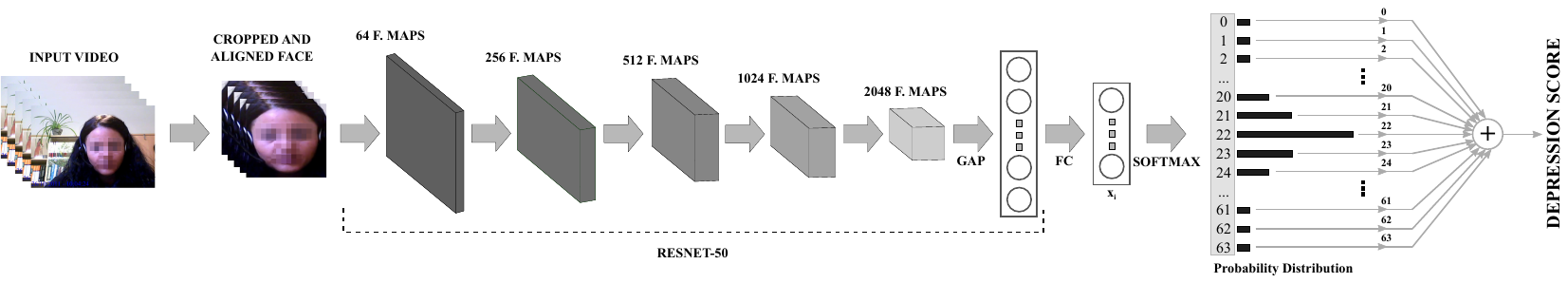}}
	\caption{The proposed method \cite{dedepression} is to estimate the severity of depression. The videos were first processed to generate the aligned facial images. Then ResNet-50 was fine-tuned to extract the discriminative features and follows a GAP layer to pool the features. Finally, expectation loss is used to weight the performance of the proposed method. }
	\label{fig:dedepression_}	
\end{figure*}
Next, in \cite{dedepression}, they adopted a 2D-CNN and distribution learning to model the patterns of depression, using the expected loss function for predicting depression levels (see Fig. \ref{fig:dedepression_}). As revealed by extensive experiments on AVEC2013 and AVEC2014, the proposed method can surpass most state-of-the-art methods (see Table \ref{table:static images}).

\begin{figure}[h]
	\centering
	\centerline{\includegraphics[scale=0.8]{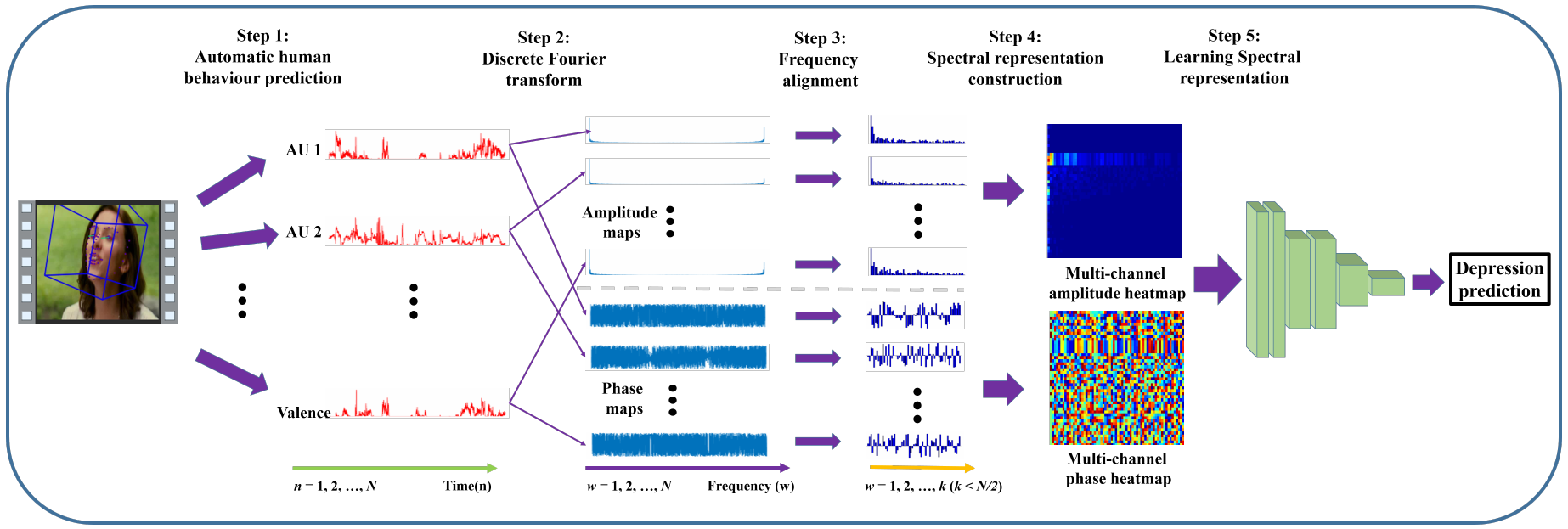}}
	\caption{The proposed method \cite{song2020spectral} to estimate the severity of depression, which can be divided into five steps: 1) Multi-channel human behaviour primitives were extracted from videos; 2) Human behaviour primitives were then transformed into spectral signals with multiple frequency patterns from all frames; 3) Owing to the symmetric of spectral signals, high-frequency pattern were removed to retain the discriminative information of human behaviours from the videos; 4) Multi-channel amplitude heatmap and multi-channel phase heatmap are constructed from spectral signals; 5) DCNN and Artificial Neural Networks (ANNs) method were used to predict the depression scales.}
	\label{fig:priminative}	
\end{figure}

 Song et al. \cite{song2020spectral} presented another novel multi-scale architecture for depression recognition (Fig. \ref{fig:priminative}). In the architecture, human behavior primitives, i.e., AUs, gaze direction and head pose, are considered either based on their occurrence (binary result) or intensity (real-valued or ordinal result) as frame-wise feature representations. Spectral heatmaps and spectral vectors are adopted to mine the multi-scale representations of expressive behavior, and then input into DCNN and Artificial Neural Networks (ANNs) for ADE. The proposed method obtained promising performance on the AVEC2013 and AVEC2014 databases (see Table \ref{table:static images}).

\begin{figure}[h]
	\centering
	\centerline{\includegraphics[scale=0.9]{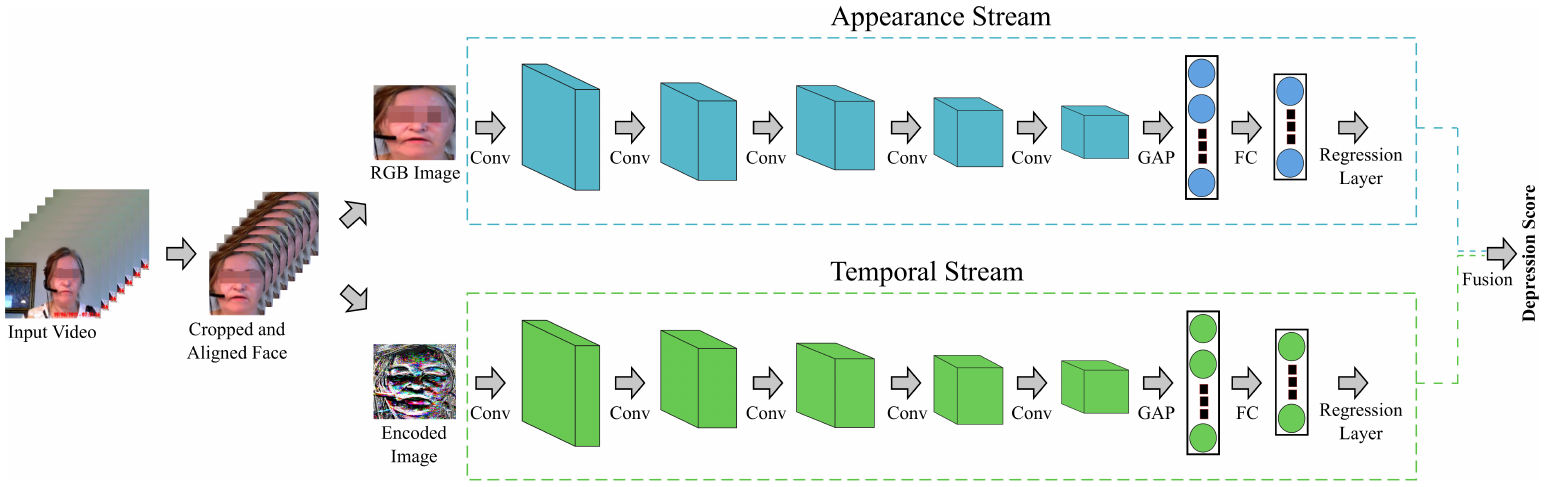}}
	\caption{This method \cite{de2020encoding} is proposed to estimate the severity of depression. The appearance stream takes the static images as input, while the temporal stream takes image sequences as input. A simple fusion method, i.e., average pooling, was used to fuse the outputs of the two networks for the \ac{ADE} task.}
	\label{fig:Two_steam}	
\end{figure}

In addition, the studies based on the AVEC2013 and AVEC2014 databases \cite{de2020encoding} introduced a two stream DCNN framework to learn the patterns from RGB images and encoded images of video clips, whose architecture is shown in Fig. \ref{fig:Two_steam}. The appearance stream takes the static images as input, while the temporal stream takes image sequences as input. The mean squared error function is used to address the regression issue. A simple fusion method, i.e., average pooling, is used to fuse the outputs of the two networks for the \ac{ADE} task.

To summarize works \cite{zhu2017automated,zhou2018visually,dedepression,song2020spectral,de2020encoding}, one can note that they hold the following general characteristics: 1) They leverage the large scale database (e.g., CASIA, VGG, etc.) to pre-train the deep models by using the deep architectures (e.g., GoogleNet, VGG, ResNet, etc.). 2) They improve the performance of the deep models by fine-tuning them on depression databases, e.g., AVEC2013 and AVEC2014, etc. 3) Moreover, some works try to improve the performance of depression recognition via designing a particular loss function for the \ac{ADE} task. 

\begin{figure}[h]
	\centering
	\centerline{\includegraphics[scale=0.5]{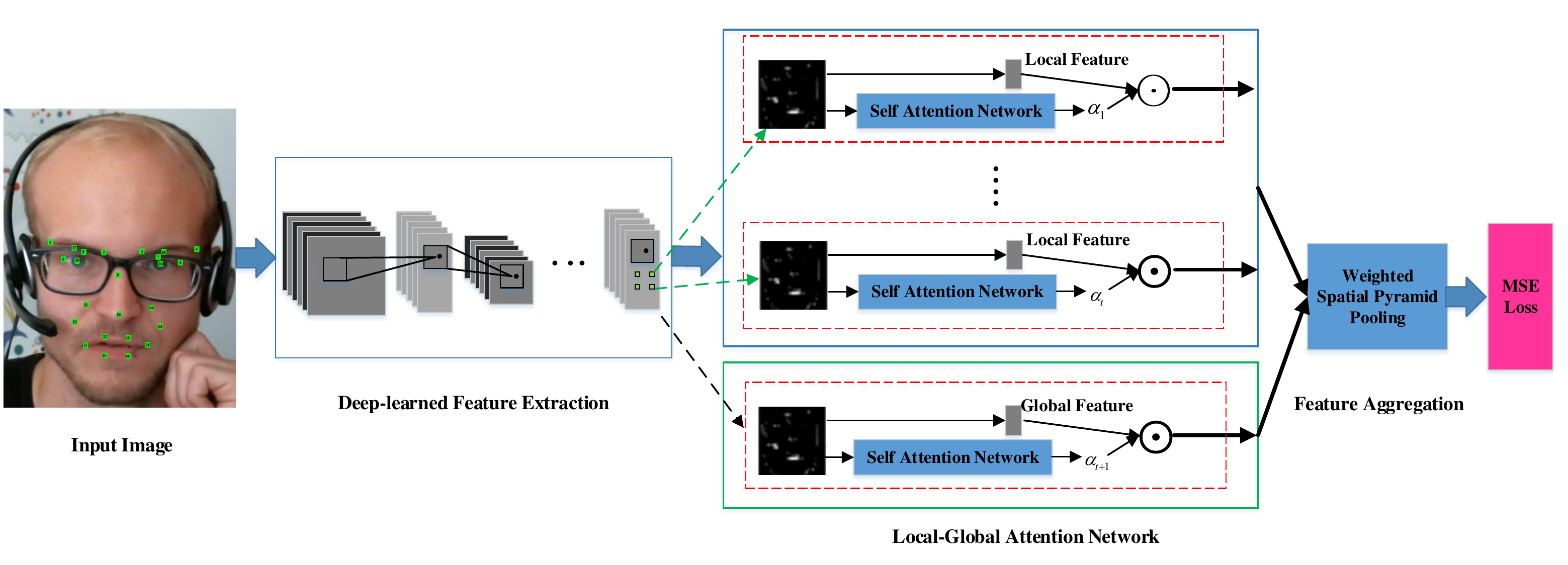}}
	\caption{The proposed method DLGA-CNN for \ac{ADE} \cite{he422automatic}. The facial image is obtained by OpenFace toolkit \cite{baltruvsaitis2016openface}. Then a typical DCNN is designed for feature representation to generate a discriminative the feature maps. To extract informative features, the local and global self-attention networks are designed. To obtain scale-invariant feature representations over multi-scale feature maps, WSPP is used. In addition, two fully connected layers and a Mean Square Error (MSE) loss layer are adopted for ADE. }
	\label{fig:he_attention}	
\end{figure}

Interestingly, He et al. \cite{he422automatic} introduced a novel network that combined a 2D-CNN networks and attention mechanism for depression recognition.
In this method, the authors proposed an integrated architecture – Deep Local-Global Attention Convolutional Neural Network (DLGA-CNN) for ADE, which utilizes DCNN with attention mechanism, and weighted spatial pyramid pooling (WSPP) to model a global feature. Two branches are designed: Local Attention-based CNN (LA-CNN) concentrates on the local patches, while Global Attention-based CNN (GA-CNN) models the global features from the entire facial region. In order to learn the complementary patterns from the two branches, Local–Global Attention-based CNN (LGA-CNN) is introduced. After the aggregation of features, WSPP is adopted to extract the depression representations. More importantly, compared to the previous methods, the proposed method did not leverage a large-scale database to pre-train the deep model, but rather is considered as an end-to-end scheme for \ac{ADE} (see Fig. \ref{fig:he_attention}).   

\begin{table*}[h]
	\centering
	\caption{Performance summary of reviewed approaches for depression recognition based on static images of the most widely assessed databases. Note that the list results of DAIC-WOZ and E-DAIC contain text features for \ac{ADE}.}
	\label{table:static images}
	\begin{tabular}{|c|c|c|c|c|c|c|}
		\hline
		\multirow{2}{*}{Modality}&\multirow{2}{*}{Datasets} & \multirow{2}{*}{Methods} & \multirow{2}{*}{Network Type} & \multirow{2}{*}{Preprocessing}  & \multicolumn{2}{c|}{Test} \\ \cline{6-7} 
		&	&                          &                               &                                      & RMSE         & MAE        \\ \hline
		
		\multirow{13}{*}{  Audio  } 
		
		&\multirow{5}{*}{AVEC2013} & He et al. 2018 \cite{he2018automated}                         & DCNN                              &  DFT                              &   10.00           &    8.20        \\ \cline{3-7} 

		&&  Niu et al. 2019 \cite{niu2019automatic}                        &  DCNN, LSTM                             &  --                               &   9.79           &   7.48         \\ \cline{3-7} 
		&&  Zhao et al. 2020 \cite{zhao2020hybrid}                       & DCNN                              &    --                             &     9.65          &  7.38          \\ \cline{3-7} 
		&&  Niu et al. 2021 \cite{NIU2021}                        &  DCNN                           &  --                               &   8.32           &   6.26         \\ \cline{3-7} 
		&&  Dong et al. 2021 \cite{dong2021hierarchical}                        &  DCNN                           &  --                               &   8.73           &  7.31         \\ \cline{2-7} 
		
		&\multirow{5}{*}{AVEC2014} &   He et al. 2018 \cite{he2018automated}                       &  DCNN                              &   DFT                             &  9.99            & 8.19           \\ \cline{3-7} 

		&&  Niu et al. 2019 \cite{niu2019automatic}                        &   DCNN, LSTM                            &   --                             &   9.66           &  8.02          \\ \cline{3-7}
		&& Zhao et al. 2020 \cite{zhao2020hybrid}                         &   DCNN                           &   --                             &   9.57          & 7.94           \\ \cline{3-7} 
		&&  Niu et al. 2021 \cite{NIU2021}                        &  DCNN                           &  --                               &   9.25           &   7.49         \\ \cline{3-7} 
		&&  Dong et al. 2021 \cite{dong2021hierarchical}                        &  DCNN                           &  --                               &   8.82           &   6.79         \\ \cline{2-7}

		&\multirow{3}{*}{DAIC-WOZ} 
		
		& Ma et al. 2016 \cite{ma2016depaudionet}                         & DCNN, LSTM                             &  --                              &  --               &    --                       \\ \cline{3-7}
		
		&&  Alhanai et al. 2018 \cite{al2018detecting}                         &    LSTM                           &   --                            &      6.50       & 5.13   \\ 	\cline{3-7}		
		&& Yang et al. 2020 \cite{yang2020feature}                         &     DCGAN                          &     --                           &  5.52            &   4.63         \\

		\hline
		\hline
		
		\multirow{12}{*}{  \tabincell{c}{ Video  \\ (Static Images)}  } 
		&\multirow{6}{*}{AVEC2013}
		& Zhu et al. 2017 \cite{zhu2017automated}                         &  DCNN                           &  Dlib                               &  9.82           &   7.58        \\ \cline{3-7} 	
		
		&& Zhou et al. 2018 \cite{zhou2018visually}                         &  DCNN                           &  Dlib                               &  8.28           &   6.20        \\ \cline{3-7} 
		&& Melo et al. 2019 \cite{dedepression}                         &  RetNet-50                           &   MTCNN                             &  8.25          &  6.30        \\ \cline{3-7}  		
		 
		&&  Melo et al. 2020 \cite{de2020encoding}                       &   DCNN                            &  MTCNN                               &  7.97            &  5.96          \\ \cline{3-7} 
		&&  Sone et al. 2020 \cite{song2020spectral}                         &       DCNN                        &      OpenFace                         &   8.10           &  6.16          \\ \cline{3-7}
		&& He et al. 2020 \cite{he422automatic}                        &  DCNN                            &     OpenFace                       & 8.39            &    6.59        \\ \cline{2-7}

		&\multirow{6}{*}{AVEC2014}   
		
		& Zhou et al. 2018 \cite{zhou2018visually}                         &  DCNN                           &  Dlib                              &  8.39           &   6.21        \\ \cline{3-7}  
		
		&&Zhu et al. 2017 \cite{zhu2017automated}                       &  DCNN                             &  Dlib                       &   9.55         &   7.47        \\ \cline{3-7}
		&&  He et al. 2020 \cite{he422automatic}                         & DCNN                             &  OpenFace                              & 8.30              & 6.51           \\ \cline{3-7} 
		&& Melo et al. 2019 \cite{dedepression}                         &  RetNet-50                           &   MTCNN                             &  8.23          &  6.15        \\ \cline{3-7} 
		
		&&  Melo et al. 2020 \cite{de2020encoding}                       &   DCNN                            &  MTCNN                              &  7.94            &  6.20          \\ \cline{3-7} 
		&&  Sone et al. 2020 \cite{song2020spectral}                         &       DCNN                        &      OpenFace                            &   7.15           &  5.95          \\ 
		
		\hline
		\hline
		
		\multirow{14}{*}{  \tabincell{c}{ Video  \\ (Image Sequences)}  }
		
		&\multirow{6}{*}{AVEC2013}
		
		& Jazaery et al. 2018 \cite{al2018video}                         &  RNN, C3D                           &  OpenFace                              &  9.28           &   7.37        \\ \cline{3-7}  
		
		&& Melo et al. 2019 \cite{de2019combining}                         &  C3D                           &    MTCNN                           &  8.26         &  6.40         \\ \cline{3-7}  
		
		&& Azher et al. 2020 \cite{uddin2020depression}                         &  DCNN, LSTM                           &   --                             &     8.93        &  7.04          \\ \cline{3-7}  
		
		&&  Melo et al. 2020 \cite{de2020deep}                       &  3D-CNN                           &  MTCNN                                     &   7.90           &   5.98        \\ \cline{3-7} 
		
		&& He et al. 2021 \cite{he2021intelligent}                        &  DCNN                            &     OpenFace                       & 8.46            &    6.83        \\ \cline{3-7} 		
		&&  Melo et al. 2021 \cite{9400725}                       &  MDN                           &  MTCNN                                           &   7.55            &  6.24      \\ \cline{2-7}

		&\multirow{6}{*}{AVEC2014}   
		
		& Jazaery et al. 2018 \cite{al2018video}                         &  RNN, C3D                           &  OpenFace                              &  9.20           &   7.22        \\ \cline{3-7}  
		&& Melo et al. 2019 \cite{de2019combining}                         &  C3D                           &    MTCNN                           &  8.31         &  6.59        \\ \cline{3-7} 
		
		&& Azher et al. 2020 \cite{uddin2020depression}                         &  DCNN, LSTM                           &   --                              &   8.78         &  6.86          \\ \cline{3-7}  
		
		&&  Melo et al. 2020 \cite{de2020deep}                       &  3D-CNN                           &  MTCNN                                           &   7.61            &  5.82      \\ \cline{3-7}
		
		&& He et al. 2021 \cite{he2021intelligent}                        &  DCNN                            &     OpenFace                       & 8.42            &    6.78        \\ \cline{3-7} 
		&&  Melo et al. 2021 \cite{9400725}                       &  MDN                          &  MTCNN                                           &   7.65            &  6.06      \\ \cline{2-7}

		&\multirow{2}{*}{DAIC-WOZ} 
		&  Song et al. 2018 \cite{song2018human}                       &      DCNN                         &                                &    5.84             &     4.37                      \\ 
		\cline{3-7} 
		&&  Du et al. 2019 \cite{du2019encoding}                       &      DCNN                         &                                &    5.78            &     4.61                      \\

		\hline
		\hline

		\multirow{12}{*}{  Multi-modal  }&
		\multirow{1}{*}{AVEC2013} 
		& Niu et al. 2020 \cite{niu2020multimodal}                         &  DCNN                             &    Dlib, STFT                                       &    8.16           &   6.14         \\ \cline{2-7} 

		&\multirow{2}{*}{AVEC2014} &  Niu et al. 2020 \cite{niu2020multimodal}                        &  DCNN                             &     Dlib, STFT                                       &   7.03           &   5.21          \\ \cline{3-7} 
		&&  Jan et al. 2018 \cite{jan2017artificial}                        &  DCNN                             &     --                           & 7.43             &  6.14          \\ \cline{2-7}
		&\multirow{6}{*}{DAIC-WOZ} & Yang et al. 2017 \cite{yang2017DCNN}                         &   DCNN, DNN                            &   --                            &   6.34            & 5.38           \\ \cline{3-7} 
		&& Yang et al. 2018 \cite{yang2018integrating}                         &DCNN, DNN                               &    --                             &   6.34            & 5.39           \\ \cline{3-7} 
		
		&& Zhao et al. 2020 \cite{8910358}                         & BLSTM                              &    --                              & 5.51             & 4.20           \\ \cline{3-7} 
		&&Yang et al. 2017 \cite{yang2017multimodal}                          & DCNN                              &  --                                & 5.97             &  5.16          \\ \cline{3-7} 
		&& Yang et al. 2017 \cite{yang2017hybrid}                         &  DCNN                             &  --                                           &  5.40            &  4.35          \\ \cline{2-7}
		
		&\multirow{4}{*}{E-DAIC}		& Shi et al. 2019 \cite{yin2019multi}                         & LSTM                              & --                             & 5.50                 &      --                    \\ \cline{3-7}
		&&Makiuchi et al. 2019 \cite{rodrigues2019multimodal}&DCNN, LSTM&--&6.11&--\\ \cline{3-7}
		&&Fan et al. 2019 \cite{fan2019multi}&DCNN&--&5.91&4.39\\ \cline{3-7}		
		&&Zhang et al. 2019 \cite{zhang2019evaluating} &Random Forest&--&6.85&5.84 \\ \hline

	\end{tabular}
\end{table*}

Table \ref{table:static images} suggests that most studies using \ac{DL} until now have adopted the DCNN architecture to predict the depression scale. In addition, attention mechanism \cite{vaswani2017attention}, has also been utilized in depression recognition \cite{he422automatic}. As for pre-processing, the researchers have mainly used MTCNN, OpenFace, Dlib toolkits to detect and crop the facial region to lay a solid foundation for depression detection.

\subsubsection{Deep ADE Networks for Image Sequences}\label{sec:Video 3D}

\begin{figure}[h]
	\centering
	\centerline{\includegraphics[scale=0.75]{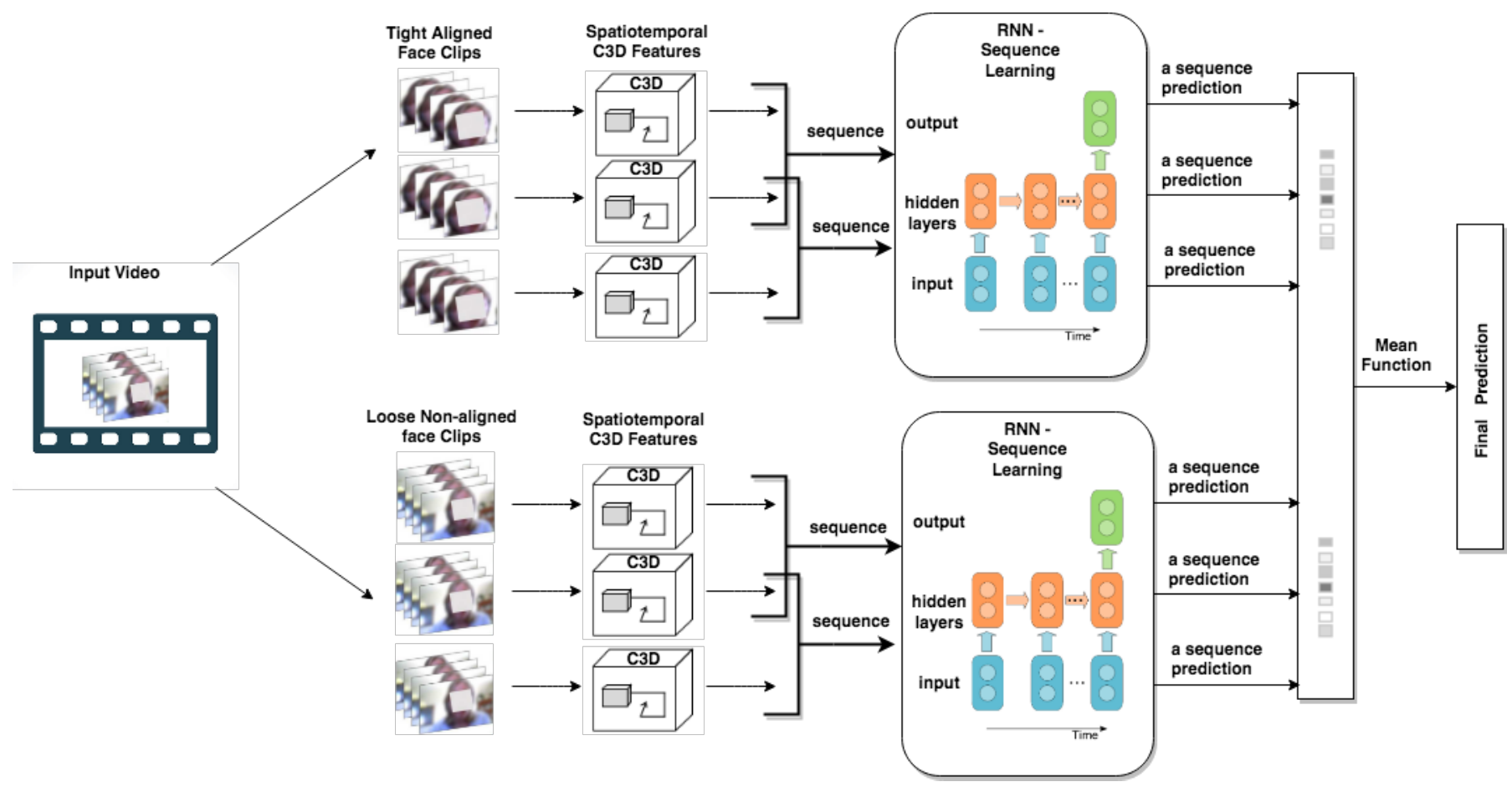}}
	\caption{The pipeline of the framework for estimating the scales of depression with Deep C3D and RNN from videos proposed in \cite{al2018video}. Discriminative features are extracted at two different scales. C3D Tight-Face model learns a tight (i.e. high-resolution) feature representation, while the C3D Loose-Face model is trained on larger face regions to learn global features. An RNN is adopted to model the temporal features based on outputs of the C3D Tight-Face and C3D Loose-Face models. Finally, a mean operation was used to generate the predictions.}
	\label{fig:RNN_C3D}	
\end{figure}

Though discriminative patterns based on single image features have been widely adopted in ADE tasks, and have obtained promising performance, these works nonetheless neglect the temporal information that could be useful for the \ac{ADE} task. To resolve this Jazery et al. \cite{al2018video} proposed using C3D and RNN to extract spatiotemporal features in two different scales from the video clips for depression recognition. The proposed framework consists of the two components, i.e., loose and tight scale feature extraction components, which use fine-tuning of deep models and temporal feature aggregation. The C3D Tight-Face model is utilized for learning of a tight (i.e. high-resolution) features, while the C3D Loose-Face model is trained on larger face regions to learn the global features. Then an RNN is adopted to model the temporal features learned by the C3D Tight-Face and C3D Loose-Face models. Finally, a mean operation is utilized for prediction. The main contribution of this work \cite{al2018video} is the temporal framework that learn facial features on different scales. Furthermore, different feature aggregation stages can combine the features from different scales, which can benefit depression scale prediction (see Fig. \ref{fig:RNN_C3D}). 

\begin{figure}[h]
	\centering
	\centerline{\includegraphics[scale=1.0]{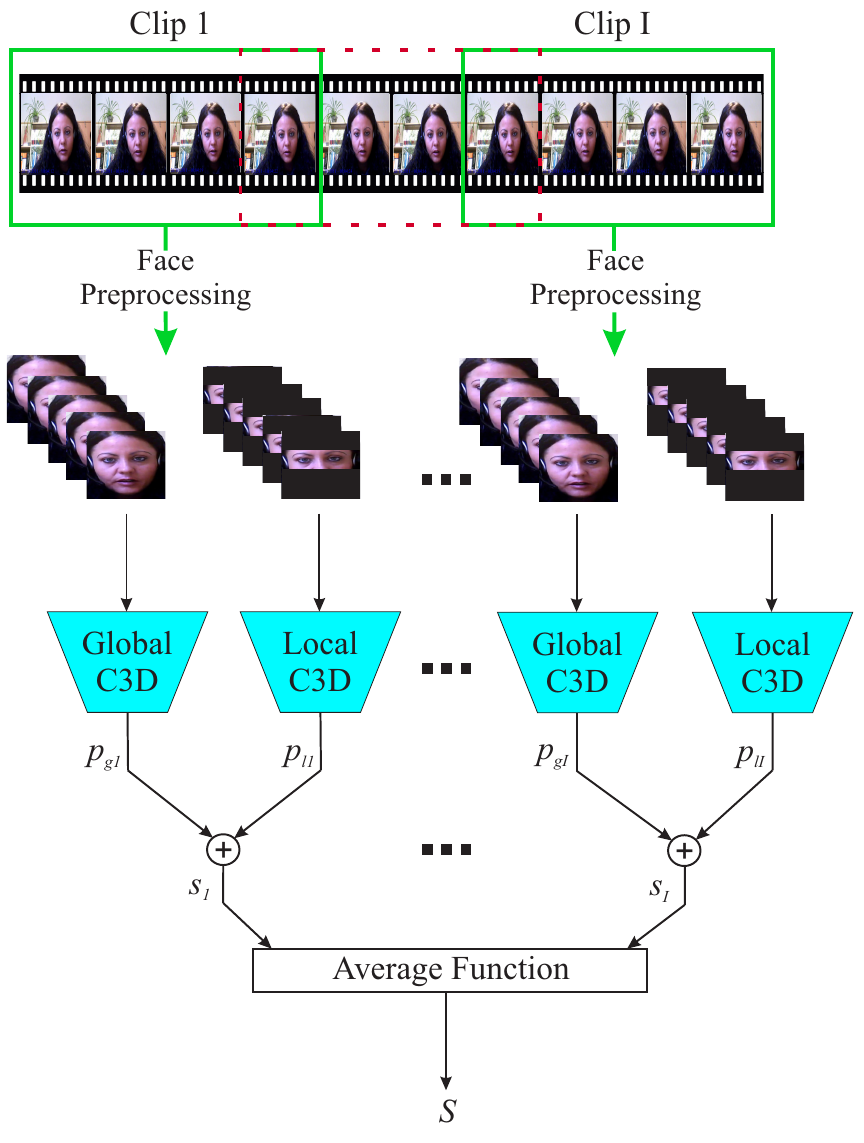}}
	\caption{The pipeline of the proposed scheme for ADE with local and global C3D from videos \cite{de2020encoding}. The video clips were pro-processed by the MTCNN toolkit. Then the two discriminative features at different scales based on C3D were extracted and concatenated. After that, the average function was used to pool the final features for predicting the depression scores. }
	\label{fig:de2020encoding}	
\end{figure}

Subsequently, Melo et al. \cite{de2020encoding} proposed a combination of different C3D architectures to learn Spatio-temporal patterns from full-face and local regions, and further combined them with 3D Global Average Pooling (3D-GAP) for predicting depression. The local C3D architecture learns discriminative information of the eye region, while the global C3D architecture focuses on learning the spatio-temporal patterns based on the whole facial region. In addition, 3D-GAP is also used to aggregate the spatio-temporal features from the last convolutional layer (see Fig. \ref{fig:de2020encoding}). The proposed method was assessed on the AVEC2013 and AVEC2014 databases, and it obtained an improved performance with the RMSEs of 8.26 and 8.31 as compared with the state-of-the-art methods, respectively (see Table \ref{table:static images}).

Uddin et al. \cite{uddin2020depression} used LSTM to model the sequence information from video data. Moreover, deep facial expression features were extracted by a deep CNN and then pooled by a Temporal Median Pooling (TMP) technology to feed the LSTM module for ADE. Various experiments were performed on the two datasets (AVEC2013 and AVEC2014), indicating that the proposed methodology surpasses most existing methods (see Table \ref{table:static images}). The contribution of work \cite{uddin2020depression} is that a Volume Local Directional Number (VLDN) dynamic feature is designed to model the trivial emotions from facial regions. 
\begin{figure}[h]
	\centering
	\centerline{\includegraphics[scale=0.9]{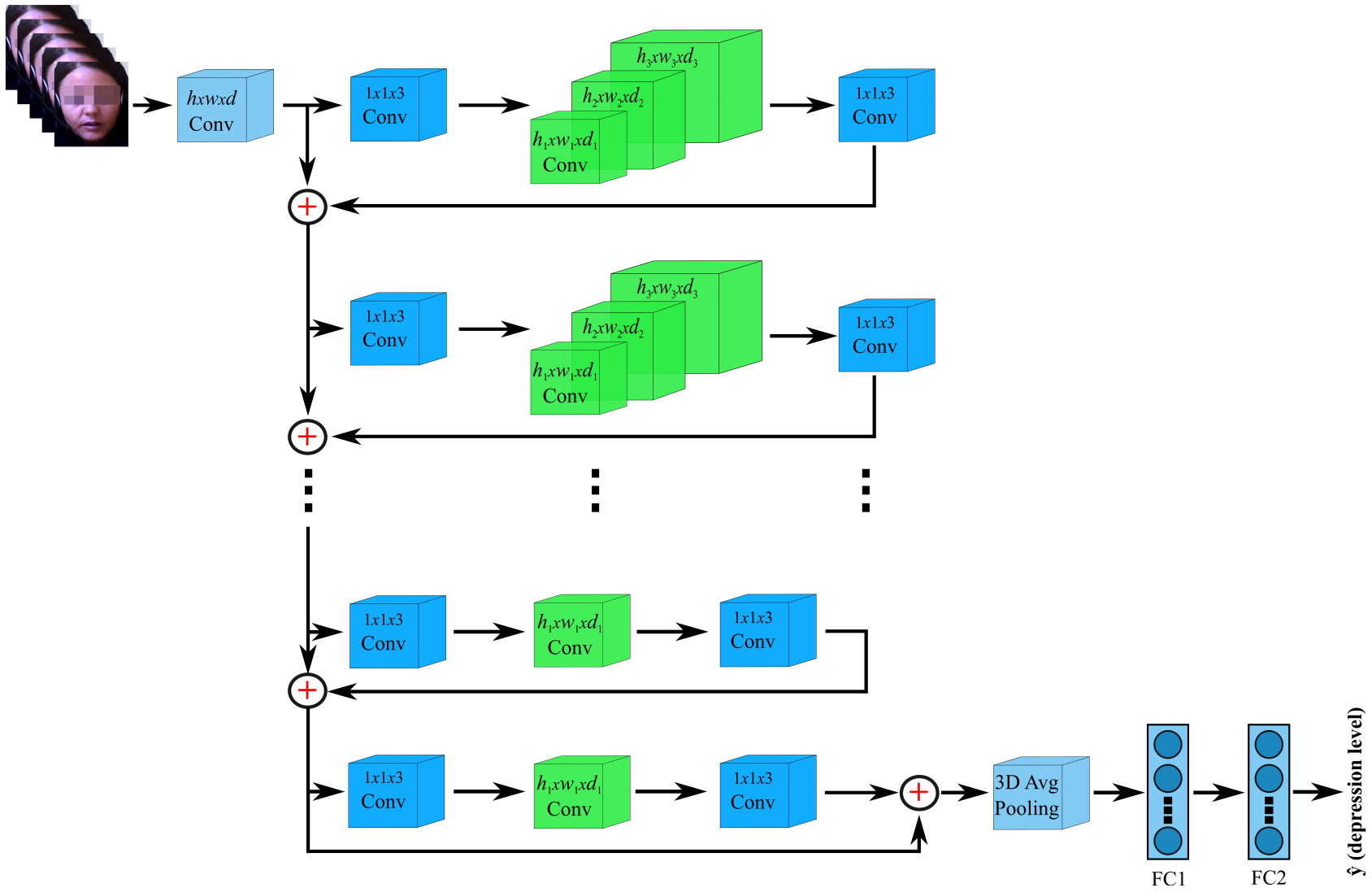}}
	\caption{The illustration of the proposed MSN for \ac{ADE} \cite{de2020deep}. In the framework, several parallel convolutional layers are used to learn considerable spatio-temporal variations from facial expressions. The model adopts several receptive fields to capture the multi-scale pattern of depression for ADE.}
	\label{fig:MSN}	
\end{figure}
In \cite{de2020deep}, a novel 3D framework, the multi-scale spatiotemporal network (MSN), was developed to learn the characteristic information of the video clips. In the work, several parallel convolutional layers were used to learn considerable spatio-temporal variations from facial expressions. The model adopts several receptive fields to maximize the exploitation of distinct spatial areas from facial region for ADE (see Fig. \ref{fig:MSN}).

\begin{figure}[h!]
	\centering
	\centerline{\includegraphics[scale=0.52]{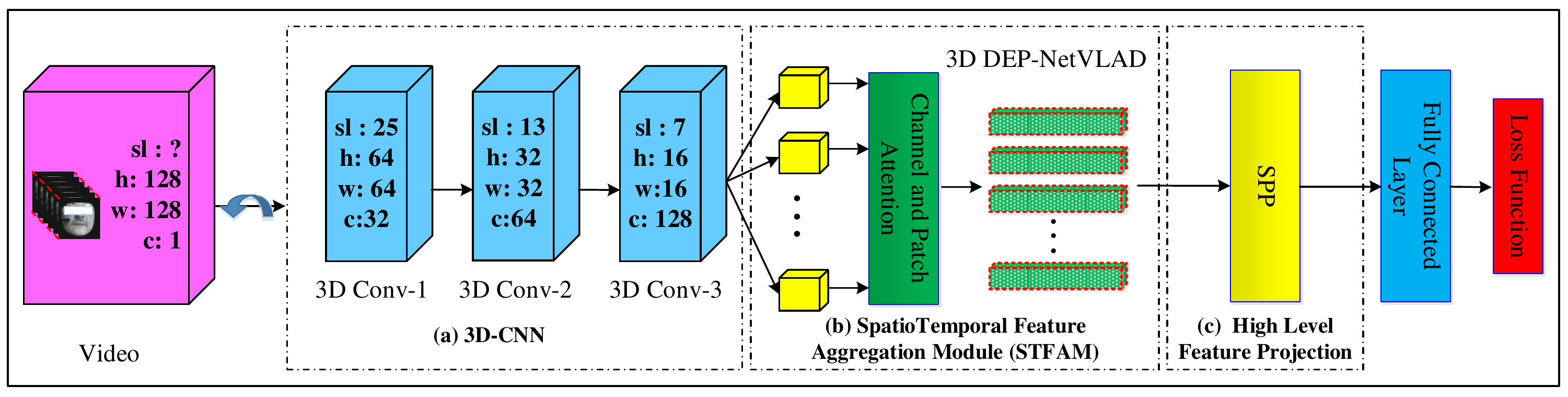}}
	\caption{The illustration of the proposed architecture for \ac{ADE} \cite{he2021intelligent}. In the framework, which consists of the following steps. The first step is that facial images are cropped and aligned by the OpenFace toolkit. The second stage is that 3D-CNN is adopted to extract the local and spatial-temporal feature representations related to the symptoms of depression. The third stage is that Spatio-temporal Feature Aggregation Module (STFAM) is used to aggregate the discriminative features over the local features. An SPP layer is adopted to represent multi-scales on top of the output of STFAM. Finally, a fully connected layer and MSE loss function is used to predict the final \ac{BDI}-II score. }
	\label{fig:3D-CNN}	
\end{figure}

In 2021, several works \cite{he2021intelligent,9400725} have proposed  to predict the severity of the depression. In \cite{he2021intelligent}, the authors proposed an end-to-end intelligent system to generate discriminative representations from the entire video clip. Specifically, a 3D-CNN combined with a Spatiotemporal Feature Aggregation Module (STFAM) is trained from scratch on AVEC2013 and AVEC2014 data, which can learn the informative patterns of depression. In the STFAM, channel and spatial attention mechanism as well as an aggregation method, namely 3D DEP-NetVLAD, are integrated to capture the compact characteristic based on the feature maps. Case studies are introduced to describe the applicability of the proposed intelligent system for ADE (see Fig. \ref{fig:3D-CNN}).

\begin{figure}[h!]
	\centering
	\centerline{\includegraphics[scale=1]{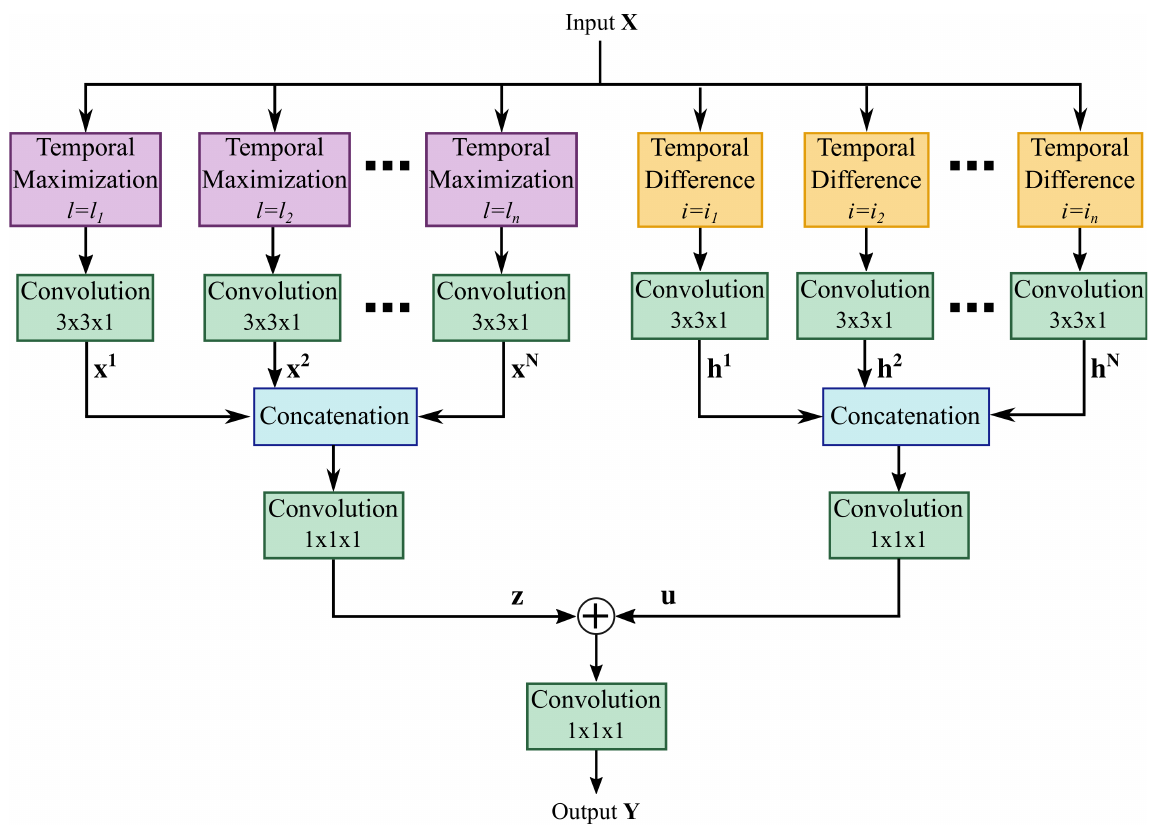}}
	\caption{A detailed illustration of the MDN module \cite{9400725}. The module consists of two blocks. The left block is a maximization block to model spatio-temporal patterns. $\textbf{X} \in \mathbb{R}^{N \times T \times H \times W \times C} $ is the feature map, in where N, T, H,  W and C represents the the batch size, temporal depth, height, width, and the number of channels, respectively. ${l_1},{l_2},....,{l_N}$ represent the branches of the maximization block. $x_i$ represents the output of $i$-th branch. $z = {\rm \mathcal{H}}\{ \bigcup\limits_{n = 1}^N {{x^n}} \}$ is the output of the maximization block. The temporal difference block (right) learns spatio-temporal variations. ${i_1},{i_2},....,{i_N}$ represent the branches of difference block. $h_i$ represents the output of $i$-th branch. $u = {\rm \mathcal{H}}\{ \bigcup\limits_{n = 1}^N {{h^n}} \}$ is the output of the difference block. Finally, the two blocks are combined to obtain the final output. } 
	\label{fig:MDN_1}	
\end{figure}

\begin{figure}[h!]
	\centering
	\centerline{\includegraphics[scale=0.83]{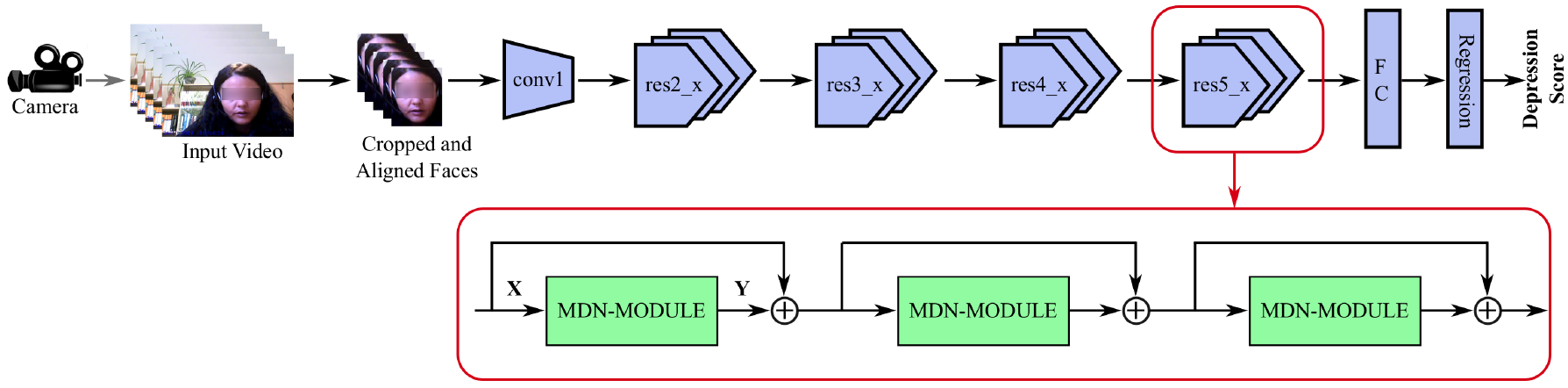}}
	\caption{Illustration of the MDN architecture \cite{9400725}. Firstly, facial images are cropped and aligned by MTCNN. Secondly, 3D residual networks \cite{hara2018can} are pre-trained on VGGFace2 dataset for image classification. Thirdly, the 3D residual deep model combined with the MDN module is fine-tuned on AVEC2013 and AVEC2014 databases to compute the BDI-II scores.} 
	\label{fig:MDN_2}	
\end{figure}

In \cite{9400725}, a new \ac{DL} architecture named Maximization and
Differentiation Network (MDN) is proposed to model the variations of facial expressions closely related to depression. The MDN is designed without 3D convolutions, and it exploits discriminative temporal patterns learned by two different blocks that model either smooth or sudden facial variations. Finally, they designed the models with 100 and 152 layers and validated the deep models on the AVEC2013 and AVEC2014 databases (see Fig. \ref{fig:MDN_1}, Fig. \ref{fig:MDN_2}). The proposed model obtains competitive RMSEs of 7.55 and 7.65 on the AVEC2013 and AVEC2014 databases, respectively (see Table \ref{table:static images}).

Based on the works mentioned in this section (see Table \ref{table:static images}), we can make the conclusions as follows:
\begin{enumerate}
	\item In comparison with static features, image sequences can capture short-term and long-term Spatio-temporal information from videos. This can improve training of a deep discriminative model for ADE task.
	\item From the perspective of training, most of the works include pre-training and fine-tuning stages for \ac{ADE}. To date, there does not exist an end-to-end scheme for \ac{ADE} from image sequences. 
	\item Summarizing the results in this section (Section \ref{sec:Video 3D}), most of the works obtained a comparable performance. So far, the method published in \cite{de2020deep} obtains the best result with RSME of 7.55 on AVEC2013, and the method from \cite{9400725} gets the best result with RMSE of 7.61 on AVEC2014, respectively.	
\end{enumerate}

\subsection{Deep ADE Networks for Multi-modal Fusion}\label{sec:Multi-modal Fusion}

\begin{figure}[h]
	\centering
	\centerline{\includegraphics[scale=0.75]{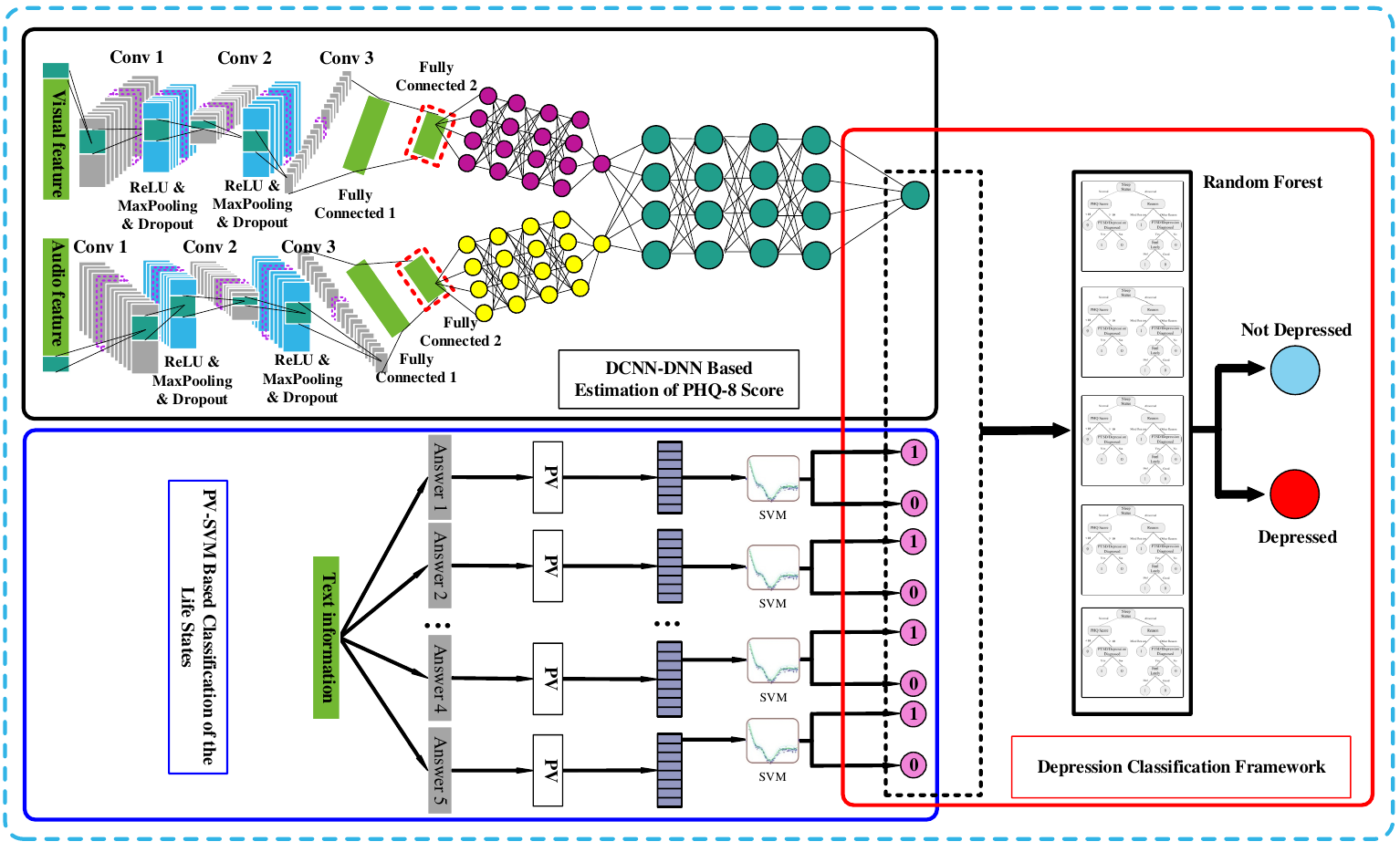}}
	\caption{The integration of shallow and deep architecture for multi-modal depression prediction and classification \cite{yang2018integrating}. The method includes three modules: 1) The audiovisual DCNN-DNN prediction module. DCNNs learn high-level features from hand-crafted audio/video features from audiovisual signals. Then PHQ-8 scores were predicted by inputting the high-level features to a DNN, and then fed to a DNN to get the final prediction (surrounded by the black rectangle). 2) The classification module. PV-SVM model is adopted to check the presence or absence of the psychoanalytic symptoms, e.g., sleeping disorder and feelings (the blue rectangle) . 3) The depression classification module. To get the final classification results based on the results of 1) and 2), a random forest method was adopted to classify the participants as the healthy controls and depressed subjects (the red rectangle). }
	\label{fig:yang_DCNN}	
\end{figure}

Apart from the above-mentioned single modalities audio (see Section \ref{sec:Audio Modality}) and video (see Section \ref{sec:Video Modality}), multi-modal fusion methods can enhance the performance of depression prediction. A combination of DCNN and DNN methods was proposed in 2017 for \ac{ADE} \cite{yang2017hybrid}, using different models to combine audiovisual features and textual inputs from a transcript. For each respective single modality, hand-crafted features were fed into a DCNN to model global-scale features, then input into a DNN to assess the PHQ-8 scores. To promote the performance of depression recognition, a multi-modal fusion scheme was formulated. Subsequently, the three single models (audio, visual, text) were fused together and input into a DNN to predict the severity of depression defined by the PHQ-8 depression scale. Moreover, Paragraph Vector (PV) was proposed to learn the distributed representations for the text descriptors. Besides, a novel video feature was proposed, i.e., Histogram of Displacement Range (HDR), capable of learning the displacements and speed of facial landmarks. Experiments were performed on the AVEC2017 challenge. It obtained comparable performance, with the RMSE of 5.97 and the MAE of 5.16 on the test set (see Table \ref{table:static images}). In \cite{yang2017hybrid,yang2017multimodal}, a hybrid depression recognition framework based on audiovisual and text descriptors was proposed. In the framework, DCNN and DNN were first used to classify depressed subjects and healthy controls. In the studies \cite{yang2017DCNN,yang2018integrating}, the method mentioned in \cite{yang2017multimodal} was also adopted to predict the severity of depression, with promising performance (see Fig. \ref{fig:yang_DCNN}). 

\begin{figure}[h]
	\centering
	\centerline{\includegraphics[scale=0.75]{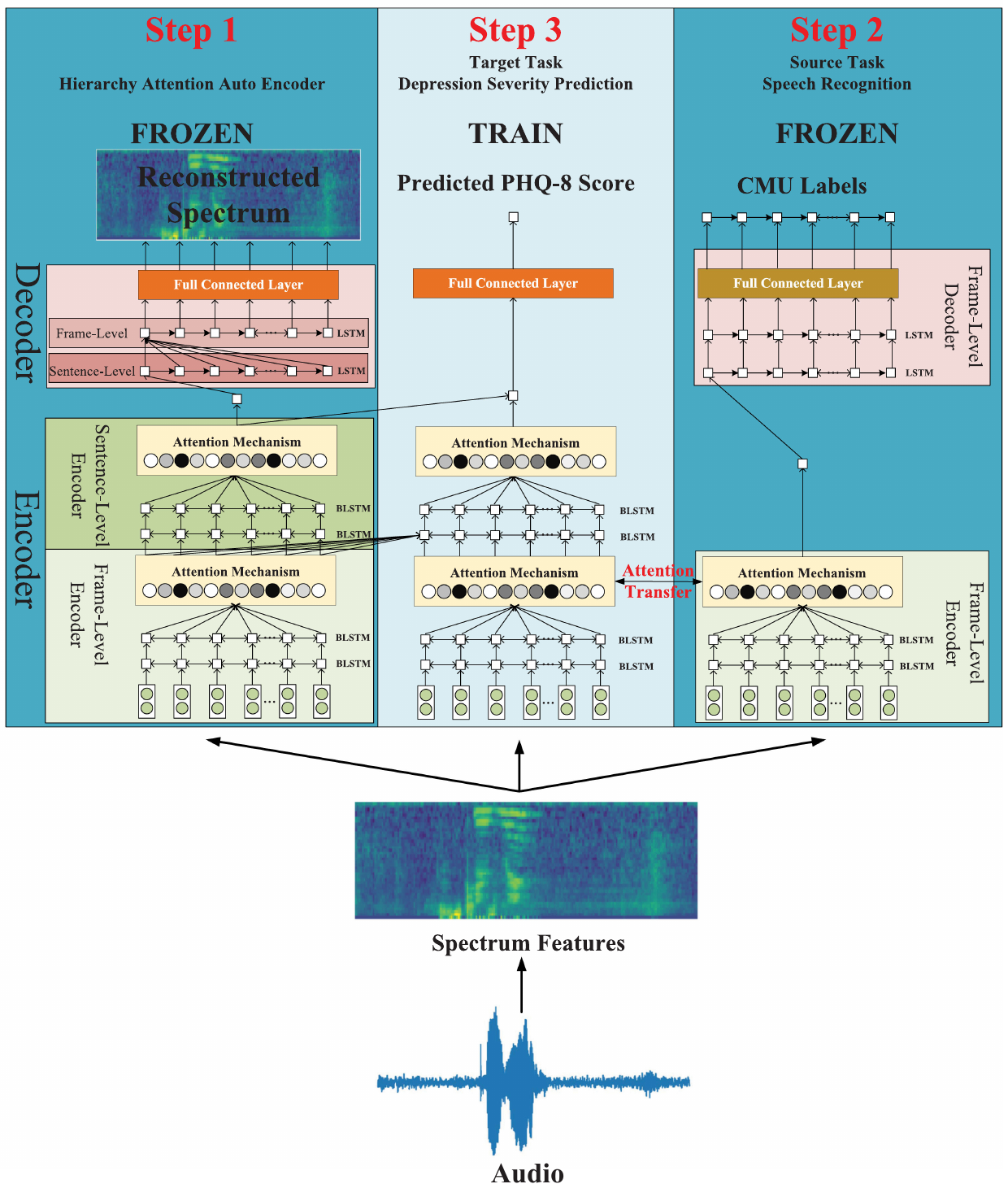}}
	\caption{A combination of hierarchical attention and auto-encoder architecture was proposed for depression recognition \cite{8910358}. First, a hierarchical attention auto-encoder network was trained based on a bottom-up scheme, in which an encoder with the attention mechanism generates a sentence-level encoder to construct the hidden vectors of the entire context. Then the vectors are performed at sentence-level attention to generate a latent representation of the clinical interview. Second, the representation was input into a decoder to re-capture the input feature. Third, the representations of the sentence are fedinto Bi-LSTM to assess the PHQ-8 scores. After that, the parameters of the auto-encoders were frozen. Then the obtained model captures its attentions by a speech recognition task, and transformed into the hierarchical depression detection system for ADE. }
	\label{fig:Fig7}	
\end{figure}

Meanwhile, a novel Bipolar Disorder Corpus was released for academic research \cite{cciftcci2018turkish} and then used for the AVEC2018 Bipolar Disorder Sub-Challenge. Based on the AVEC2018 database, in \cite{yang2018bipolar}, a novel architecture fusing a DNN and Random Forest was proposed for bipolar depression analysis. In \cite{du2018bipolar}, to address bipolar disorder (BD) with irregular variations among different episodes, a new architecture, i.e., IncepLSTM, was designed, capable of combining Inception module and LSTM of feature sequence with learning multi-scale temporal patterns for BD analysis. Experiments were performed on the AVEC2018 dataset, demonstrating the efficiency of the proposed method. In addition, other works also adopted conventional machine learning methods for BD recognition \cite{syed2018automated,xing2018multi}. However, the AVEC2018 database has not been extensively employed by researchers from the affective computing community so far. Notably, \cite{8910358} presented a novel method integrating unsupervised learning, transfer learning, and hierarchical attention from speech to assess the depression scale (see Fig. \ref{fig:Fig7}). The proposed method was evaluated on the AVEC2017 depression challenge, and RMSE and MAE are 5.51 and 4.20, respectively (see Table \ref{table:static images}).

In addition, two feature sets were introduced to determine the duration of sequential landmarks. As reported by extensive experiments evaluated on DAIC-WOZ and SH2 databases, the model can learn the patterns effectively as compared with the previous speech-based depression recognition methods \cite{8861018}. 

\begin{figure}[h]
	\centering
	\centerline{\includegraphics[scale=0.65]{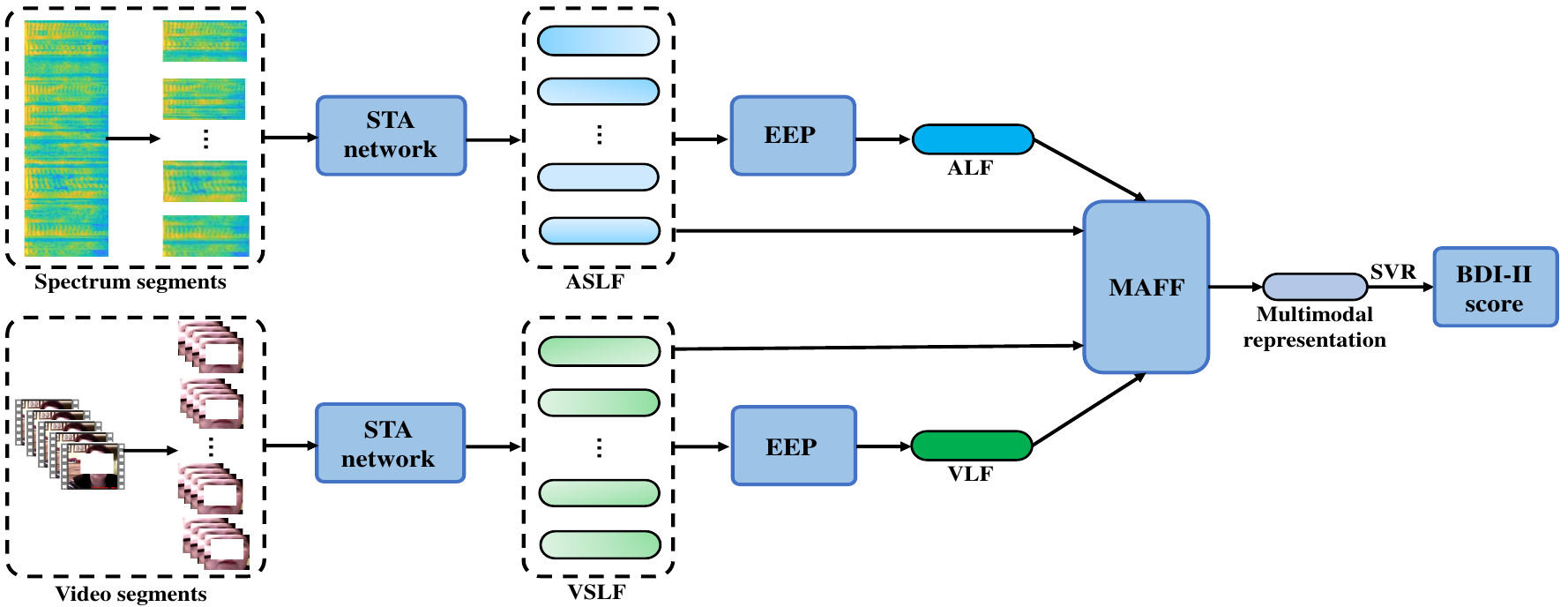}}
	\caption{The pipeline of the multi-modal spatiotemporal architecture for deep depression recognition. The approach first inputs spectrogram/video segments into the Spatio-Temporal Attention (STA) network and then uses the features of the last fully connected layers as Audio Segment-Level Feature (ASLF) and Video Segment-Level Feature (VSLF), respectively. Subsequently, Eigen Evolution Pooling (EEP) approach is used for pooling the ASLFs and VSLFs to the ALF and VLF. Lastly, Support Vector Regression (SVR) is used for predicting the \ac{BDI}-II scores \cite{niu2020multimodal}.}
	\label{fig:Niu_multimodal}	
\end{figure}

To learn the auxiliary information between audio and video cues, a new Spatio-Temporal Attention (STA) architecture and a Multi-modal Attention Feature Fusion (MAFF) method was proposed to extract the multi-modal features from audiovisual cues for predicting the depression scale, i.e., \ac{BDI}-II score. The proposed method comprises 2D-CNN, 3D-CNN, and an attention mechanism to learn the deep features for depression scale prediction. Extensive experiments are carried on the AVEC2013 and AVEC2014 databases, demonstrating that the proposed deep architecture outperformed most of the existing studies \cite{niu2020multimodal} (see Fig. \ref{fig:Niu_multimodal}). To sum up, the mentioned works appeared for the image sequences for depression recognition, and they leverage mature \ac{DL} technology (e.g., DCNN, RNN, LSTM) to learn the deep discriminative patterns for depression estimation. In addition, an attention mechanism is also used to learn the salient patterns from the deep-learned features. Moreover, in the Detecting Depression with AI Sub-Challenge (DDS) of AVEC2019, several works are also focused on adopting AI technology to estimate a subject's depression scale (Table \ref{table:static images}).    

Also, the authors in \cite{al2018detecting} used the LSTM to model the interactions in audio and text features based on the AVEC2017. 

To summarize our findings (see also Table \ref{table:static images}): 

\begin{itemize}
	\item From the perspective of modality, the multi-modal fusion method yielded the optimal performance for \ac{ADE} on every database. On the AVEC2013 and AVEC2014 databases, Niu et al. \cite{niu2020multimodal} achieved the best accuracy with RMSE of 7.03, and MAE of 5.21. On the DAIC database, the method proposed in \cite{yang2017hybrid} obtained the best performance with RMSE of 5.40, and MAE of 4.35, respectively. Though multi-modal fusion yields the optimal performance for \ac{ADE}, this method is very complicated when fusing the complementary information between audio and video cues. Consequently, a wide range of researchers have focused on the video modality to learn the discriminative patterns around facial regions, which is probably explained by the success of computer vision and \ac{DL} in general. 
	
	\item From the database point of view, AVEC2013, and AVEC2014 have obtained the most attention. The reason is that the audio and video clips are contained in the AVEC2013 and AVEC2014 databases. Thus, the researchers can leverage \ac{DL} technology to learn the compact representation from the video clips. For the DAIC database, the database organizers only provide the audio data samples, limiting their use for \ac{ADE}. 
	
	\item From the perspective of \ac{DL}, the DCNN is commonly used by different works to learn the discriminative patterns from both static images and hand-crafted features. To model the sequential information from the video sequences, a 3D-CNN is also used in many works. Different variations based on 3D-CNN are also proposed to predict the severity of depression \cite{9400725}. 
	
\end{itemize}

Furthermore, since 2015, there also exist works that have not adopted \ac{DL} technology for depression estimation, (e.g., \cite{anis2018detecting,he2018automated,alghowinem2020interpretation, jayawardena2020ordinal,he2018automatic}). In particular, Sadari et al. \cite{jayawardena2020ordinal} used ordinal logistic regression for depression recognition and proposed a new way for the task. In addition, based on the database of AVEC2017, numerous methods have been proposed for depression recognition. In \cite{8682956}, the authors analyze the relationship between the ground truth and predictions when measuring the severity of depression. They design a system validated on the AVEC2017 depression database. They found that depression recognition is an ordinal problem. Also, He et al.\cite{he2018automatic} introduced a promising feature descriptor, i.e., median robust LBP-TOP (MRLBP-TOP), that can learn  patterns on different scales from image sequences. Dirichlet process FV (DPFV) has also been proposed to learn the global patterns from the segment-level features. In addition, Bipolar Depression (BD) has also aroused attention in the affective computing field. According to Table \ref{table:BD}, notably, various works have been considered for estimating the BD. Researchers still continue using DCNN, LSTM and DNN to extract the deep features to represent the severity of BD from the perspective of methods.

\begin{table*}[h]
	\centering
	\caption{Performance summary of reviewed approaches for multi-modal depression recognition on the BD databases.}
	\label{table:BD}
	\begin{tabular}{|c|c|c|c|c|c|}
		\hline
		\multirow{2}{*}{Datasets} & \multirow{2}{*}{Methods} & \multirow{2}{*}{Network Type} & \multirow{2}{*}{Preprocessing} & Dev.        & Test       \\ \cline{5-6} 
		&                          &                               &                                & \multicolumn{2}{c|}{UAR} \\ \hline
		
		\multirow{8}{*}{BD}       &  Yang et al. 2018 \cite{yang2018bipolar}                        &    DCNN                           &  Openpose                     &         & 57.41\%                     \\ \cline{2-6} 
		&   Xing et al. 2018 \cite{xing2018multi}                       &  GBDT                             &    --                    &        &  57.41\%                    \\ \cline{2-6} 
		&   Du et al. 2018 \cite{du2018bipolar}                       &    LSTM                          &   --                     &  65.1\%      &    --                  \\ \cline{2-6} 
		& Shahin et al. 2019 \cite{amiriparian2019audio}                        &  DCNN                             &  --               &   --           &  45.5\%                    \\ \cline{2-6} 
		& Sun et al. 2021 \cite{amiriparian2019audio}                         &   DCNN                            &   --                             & 93.12\%      &   --            \\ \cline{2-6} 
		&Li et al. 2019 \cite{li2019audio}&DCNN & --& 74.5\% & \\ \cline{2-6}
		&Ren et al. 2019 \cite{ren2019multi}&DCNN, DNN&--&--& 57.41\% \\ \cline{2-6} 
		&Abaei et al. 2020 \cite{abaei2020hybrid}&DCNN, LSTM&--&60.67\%&--\\ \hline
		
	\end{tabular}
\end{table*}

\section{Open Issues and Promising Directions}\label{sec:Additional}

This section introduces open challenges in \ac{ADE} and suggests promising future directions. We want to encourage the affective computing community to study together on the mentioned issues to boost \ac{ADE} development. Consequently, our goals are: (1) to promote AI-based \ac{ADE} frameworks to apply in real life, especially in hospitals, psychiatric centers, and so on. This comprises assessing and polishing prototypes for clinical applications; promoting the availability, extensibility, and capacity in non-laboratory applications. (2) To significantly promote the research of \ac{ADE} in the future. In the discussion we specifically focus on the availability of databases, the transparency of code, the collaboration of research groups, and the imbalanced distribution of training samples.  

\subsection{The Availability of Databases}\label{sec:The Availability of Databases}

Because of the sensitivity of depression data, it is difficult to gain various data for estimating the scale of depression. Hence, the availability of data is a major issue. First, as opposed to the facial expression recognition task, the availability of databases is scarce up until the present day. Given the literature review, one can note that the widely used depression databases are AVEC2013, AVEC2014, DAIC-WOZ. Notably, AVEC2014 is a subset of AVEC2013. Second, there is no multi-modal (i.e., audio, video, text, physiological signals) database to learn comprehensive depression representations for \ac{ADE}. The existing databases consist of two or three modalities. Though the DAIC database comprises three modalities (audiovisual and text), the organizer has not provided the original videos of DAIC, leading to a certain inconvenience for \ac{ADE}. Third, the limited size of the data sets limits
the research in depression prediction, especially when using \ac{DL} technologies. For instance, AVEC2013 only contains 50 samples for the training, development, test set, respectively. To address this bottlenec, effective methods to augment the limited amount of annotated data are called for. Fourth, the criteria for data collection should be standardized. At present, different organizers adopt a range of conditions, equipment, and configurations to collect the multi-modal data.          

\subsection{The Transparency of Data and Algorithms}\label{sec:The Transparency of Code}

Despite the significant advances in \ac{ADE} tasks, there still exist many aspects for improving performance in clinical environments. Nowadays, many researchers from the affective computing community may not share their algorithms using web applications (i.e., github, personal web site). As we know, the DAIC-WOZ dataset has been widely used by many researchers for \ac{ADE} in the affective computing field. However, the publisher of DAIC-WOZ considers it difficult to provide the data as raw video clips due to the sensitivity of the personal attributes related to mental disorders. Hence, we encourage the all studies to share also the raw data and not only hand-crafted features, or at least to arrange an access to the data in a secure computing environment if sharing is not feasible. 

At the very least, researchers should make their code publicly available. Accordingly, different researchers can validate the efficiency of algorithms to build a solid foundation for clinical application. For instance, feature extraction is important for ensemble \ac{ADE}. However, the current bottleneck is to know which features are suitable for \ac{ADE}. Given that the principal way of learning fetaures for depression scale prediction is to use deep learning, researchers should design a network that is most suitable for this task. Currently, no commonly accepted standard \ac{DL} architecture has been defined for \ac{ADE}.

\subsection{The Collaboration of Research Groups}\label{sec:the collaboration of research groups}

With the significant progress among different disciplines, collaboration with other disciplines is crucial for \ac{ADE}. For the topic of affective computing, relevant fields include psychology, physiology, computer science, machine learning, etc. Thus, researchers should borrow each other's strengths for promoting the advances of \ac{ADE}. For audio-based \ac{ADE}, the deep models only represent the depression scale from audios. Specific to video-based \ac{ADE}, the deep models capture patterns only from facial expressions. Notably, physiological signals also contain significant information closely related to depression estimation. Accordingly, different researchers should study together to build a multi-modal based \ac{DL} approaches for clinical application.  

\subsection{The Imbalanced Distribution of Training Samples}\label{sec:The Imbalanced Distribution of Training Samples}

\begin{figure}[h]
	\centering
	\centerline{\includegraphics[scale=0.45]{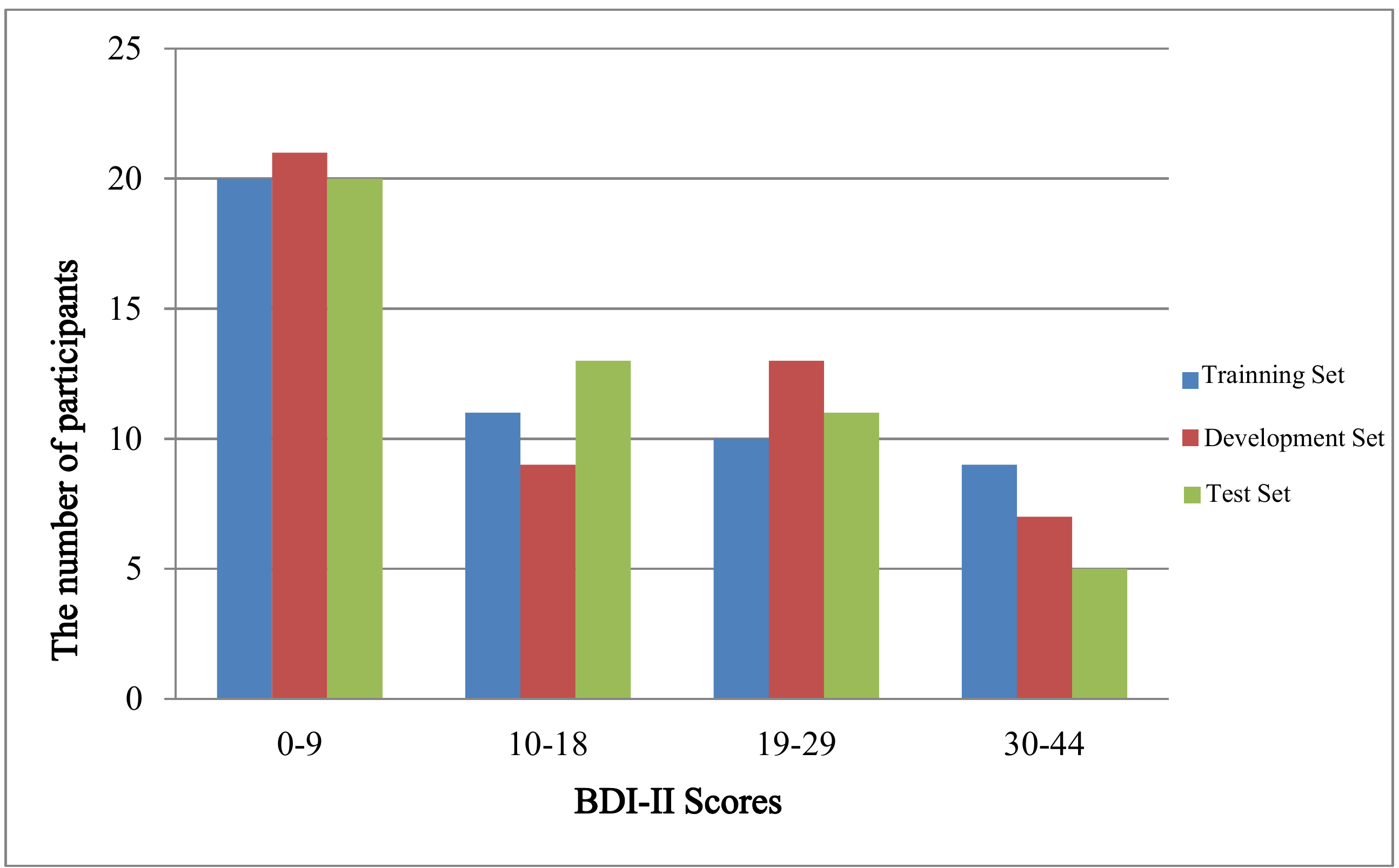}}
	\caption{The distribution of BDI-II scores in the training, development, test set of AVEC2014. The range of \ac{BDI} scores is from $0$ to $63$ (\textit{(no or minimal depression: the range from 0 to 13)}, \textit{(Mild: the range from 14 to 19)}, \textit{(Moderate: the range from 20 to 28)}, \textit{(Severe: the range from 29 to 63)}). The maximum BDI-II score of AVEC2014 is 45. The Y axis is the number of samples in the range of BDI-II. }
	\label{fig:label}	
\end{figure}

Besides the issues mentioned above, another main issue is the imbalanced distribution of training samples. This issue originates from the fact that the severity of depression is assessed by different discrete numerical values. There are two challenges in modeling the imbalanced data. Firstly, training with imbalanced data may lead to obtaining poor performance of the trained model in the minority category \cite{he2009learning}. Secondly, training the models on imbalanced data sets can markedly affect the validation/test set performance. Jeni et al. \cite{jeni2013facing} investigate the effect of skew based on the imbalanced validation set. To address the imbalance, the findings are based on several evaluation metrics, i.e., Accuracy, F1-score, and so on. To demonstrate the imbalance, consider AVEC2014 which is a popular database used by many in the affective computing field. As mentioned in Section \ref{subsec:databases}, the \ac{BDI}-II scores can be divided into four classes according to the depression scales from mild to severity, i.e., 0-9, 10-18, 19-29, 30-44. As illustrated in Fig. \ref{fig:label}, the 0-9 class has  more and the 30-44 class fewer participants compared to the other classes. Therefore, the database providers should consider the issue of data imbalance to facilitate the training of the shadow or deep models for depression analysis.
\section{Conclusions}\label{sec:Conclusions}

Along with the progress of deep learning \cite{ma2016depaudionet}, multiple works using \ac{DL} have also been proposed \ac{ADE}, with promising performance, establishing the foundations for the clinical application of \ac{ADE} systems. The present comprehensive survey of the existing \ac{ADE} methods reviews the topic from several angles while also highlighting numerous issues for further exploration. As a mental disorder, the diagnostics of depression rely on a concerted effort from multiple fields including clinical psychology, affective computing, and computer science. Based on the problems mentioned, developing automatic, objective evaluation systems is valuable for both academic research as well as clinical application.
At present, several issues remain to be addressed: 1) the ability to make a distinction between MDD and other depression types \cite{stratou2015automatic}; 2) the capacity of learning from few training samples; 3) the ability to extract discriminative features by hand-crafted and deep learning methods; 4) the ability to represent and combine the complementary information from audio and visual cues by fusion approaches. 

We conclude by emphasizing the considerable potential in the clinical scenario for assessing the severity of depression. Despite the great progress in recent years, to assist the clinical application, more work should be conducted to collect the additional data, explore a range of methods, and design implement \ac{ADE} systems for clinical use cases. 

In the future, we will resolve the following issues:
\begin{enumerate}
    \item For the small samples of the training data, on the one hand, we should encourage the data organizer to share the private data samples for ADE. On the other hand, we will try to collect a multi-modal database that includes audio, video, text, physiological signals (i.e., EEG, ECG, etc.). Therefore, different modalities can augment the data samples for training the ADE models. In addition, we would like to encourage the researchers to share their code on different platforms. 
     
    \item To extract the informative features of multi-modal cues, we will consider the attributes of individuals by the DL method. Meanwhile, we will leverage the attributes of data to extract the informative and discriminative features for ADE. In addition, we will collaborate with the researcher from interdisciplinary areas to extract the more informative feature closely related to depression. 
    
    \item To learn the complementary patterns between hand-crafted and deep-learned features, robust methods will be designed for ADE. Though deep-learned features have been proven to obtain promising ADE performance, the transnational hand-crafted features should not be ignored for ADE tasks. Therefore, we will deeply study the complementary characteristic between hand-crafted and deep-learned features to model the discriminative architecture for ADE.
    
    \item Multi-modal data are not only augment the size of data to train the modality, but also capture the discriminative patterns for ADE. To improve the performance of the multi-modal ADE, we will consider the complementary patterns among different modality and draw on the experience of researchers from different fields. All in all, we drive the ADE research into clinical application to benefit for those depressed subjects in the future. 
\end{enumerate}

\section{Acknowledgment}

This work is supported by the Scientific Research Program Funded by Shaanxi Provincial Education Department (Program No. 20JG030), the Special Construction Fund for Key Disciplines of Shaanxi Provincial Higher Education, the Shaanxi Higher Education Association Fund for the Prevention and Control of Novel Coronavirus Pneumonia (grant XGH20201), the Shaanxi Provincial Public Scientific Quality Promotion Fund for Emergency Popularization of COVID-19 (grant 2020PSL(Y)040). This work was supported by the Academy of Finland (grants 336033, 315896), Business Finland (grant 884/31/2018), and EU H2020 (grant 101016775). We would also like to thank Dr. Jonathan Cheung-Wai Chan from VUB for proof-reading the manuscript.

\bibliography{mybibfile}

\end{document}